\documentclass[12pt,a4paper]{article}

\usepackage{lmodern}
\usepackage{epsf}
\usepackage[latin1]{inputenc}
\usepackage[english]{babel}
\usepackage{amsfonts}
\usepackage{amssymb}
\usepackage{amsmath}
\usepackage{bm}
\usepackage{graphicx}

\usepackage{multirow}
\usepackage{multicol}
\usepackage{array}
\usepackage{version}
\usepackage{url}
\usepackage{lscape}
\usepackage{mathrsfs}
\usepackage[nottoc]{tocbibind} 
\usepackage{layout} 
\usepackage[textwidth=16.5cm,textheight=22cm]{geometry}
\usepackage{makecell}
\usepackage{lipsum}
\usepackage{cite}
\usepackage{float}
\usepackage{quoting}
\usepackage{authblk}
\usepackage{color}
\usepackage{enumerate}

\title{\textbf{Inflationary magnetogenesis with added helicity: constraints from non-gaussianities}}

\author[a]{Chiara Caprini\thanks{caprini@apc.in2p3.fr}}
\author[b,c]{Maria Chiara Guzzetti\thanks{mariachiara.guzzetti@pd.infn.it}}
\author[d]{Lorenzo Sorbo\thanks{sorbo@physics.umass.edu}}
\affil[a]{Laboratoire Astroparticule et Cosmologie, CNRS UMR 7164, Universit\'e Paris Diderot, 10 rue Alice Domon et L\'eonie Duquet
75013 Paris (France)}
\affil[b]{Dipartimento di Fisica e Astronomia ``G. Galilei'', Universit\`a degli Studi di Padova, via Marzolo 8, I-35131, Padova, Italy}
\affil[c]{INFN, Sezione di Padova, via Marzolo 8, I-35131, Padova, Italy}
\affil[d]{Amherst Center for Fundamental Interactions, Department of Physics, University of Massachusetts, Amherst, MA 01003}

\begin{document}

\maketitle

\abstract{In previous work~\cite{Caprini:2014mja}, two of us have proposed a  model of inflationary magnetogenesis based on a rolling auxiliary field able both to account for the magnetic fields inferred by the (non) observation of gamma-rays from blazars, and to start the galactic dynamo, without incurring in any strong coupling or strong backreaction regime. Here we evaluate the correction to the scalar spectrum and bispectrum with respect to single-field slow-roll inflation generated in that scenario. The strongest constraints on the model originate from the non-observation of a scalar bispectrum. Nevertheless, even when those constraints are taken into consideration, the scenario can successfully account for the observed magnetic fields as long as the energy scale of inflation is smaller than $10^6\div10^8$~GeV, under some conditions on the slow roll of the auxiliary scalar field.}


\def\bk{{\bf k}}
\def\bq{{\bf q}}
\def\bx{{\bf x}}
\def\bp{{\bf p}}

\section{Introduction}%

According to the non-observation of GeV gamma-ray cascades around blazars, see e.g.~\cite{Neronov:1900zz,Taylor:2011bn,Vovk:2011aa}, magnetic fields with correlation length $L$ in the intergalactic medium should have minimal amplitude of the order of $10^{-18}$~G for $L\gtrsim D_e$, with $D_e$ the electron/positron energy loss length for inverse Compton
scattering, which in this context is typically $D_e\simeq 80$ kpc \cite{Neronov:2009gh}. If on the other hand $L<D_e$, the bound changes by a factor $\sqrt{D_e/L}\,\,\Pi^{-1}(D_e/L,n_B)$, where $\Pi(D_e/L,n_B)$ is a function provided in \cite{Caprini:2015gga} and $n_B$ denotes the spectral index of the magnetic field power spectrum, see section \ref{sec:gauge}. Moreover, while there is no consensus on the origin of the magnetic fields observed in galaxies (see~\cite{Brandenburg:2004jv} and references therein), it is possible that they are the product, via the dynamo mechanism, of the amplification of a primordial ``seed'' field. Under some circumstances, depending on the correlation scale $L$ and on the magnetic field spectral index $n_B$, the intergalactic magnetic field could provide the seed for the galactic dynamo.

The most straightforward explanation for the origin of the intergalactic magnetic field, which do not seem to be associated with any cosmic structure, is that they have been formed prior to structure collapse, possibly in the pre-recombination universe. However, even such a small amplitude as required by the lower bound inferred from blazars is  challenging to account for by generation mechanisms operating in the early universe. 

Inflation has been often considered as a possible stage for the formation of magnetic seeds, because it can provide favourable conditions for magnetogenesis. The simplest way to achieve this is to postulate that the gauge field is coupled in a gauge-invariant fashion to some rolling degree of freedom during inflation\footnote{Differently motivated settings have also been proposed, see e.g.~\cite{Nandi:2017ajk,Tasinato:2014fia}}~\cite{Turner:1987bw}. The rolling field can be the inflaton field coupled to the operator $F_{\mu\nu}\,F^{\mu\nu}$~\cite{Ratra:1991bn} (which might be a pseudoscalar coupled to $F_{\mu\nu}\,\tilde{F}^{\mu\nu}$~\cite{Garretson:1992vt}) or some auxiliary field. The gauge field modes are amplified by the rolling scalar; at the end of inflation, the electric field is dissipated by the high conductivity of the universe and a magnetic field is left. If the dynamics responsible for the amplification of the gauge is parity-violating, then typically only one gauge field polarisation is amplified~\cite{Anber:2006xt}, leading to helical magnetic fields -- see also the numerical analysis of~\cite{Adshead:2016iae}. During the subsequent radiation-dominated epoch helical fields undergo a process of inverse cascade, first put forward in~\cite{Son:1998my}, that increases the comoving power of the magnetic field at large scales\footnote{Note that it has been recently discovered that energy cascades can also occur for non-helical magnetic fields \cite{Brandenburg:2014mwa,Kahniashvili:2015msa}}. The presence of a helical magnetic field of primordial origin might also be associated to the generation of the baryon asymmetry of the Universe~\cite{Anber:2015yca,Fujita:2016igl,Kamada:2016eeb,Kamada:2016cnb,Jimenez:2017cdr}, even if the mechanism is not efficient enough if CMB constraints from the production of primordial tensor modes are accounted for~\cite{Papageorgiou:2017yup}.

Successful models of inflationary magnetogenesis must avoid issues such as excessive back-reaction on the inflationary background~\cite{Martin:2007ue} or the occurrence of strong coupling~\cite{Demozzi:2009fu}. Moreover, the gauge field is a source of metric perturbations, scalar~\cite{Bonvin:2011dt} and tensor~\cite{Sorbo:2011rz} (the vector modes decay). Being the source quadratic in $A_\mu$, both scalar~\cite{Barnaby:2010vf} and tensor modes~\cite{Cook:2013xea} are in general non-gaussian. Furthermore, if only one gauge field polarisation is amplified due to parity violation, the tensor modes it leads to are chiral \cite{Sorbo:2011rz}. While the last property is welcome as a possible fingerprint of tensor modes produced by gauge fields, non-gaussianity in the curvature perturbations represents a serious issue and can lead to severe constraints on this kind of models once CMB limits on $f_{\rm NL}$ are accounted for~\cite{Barnaby:2012tk,Barnaby:2012xt}. In~\cite{Ferreira:2014zia} it was shown that, even if the scalar field to which the gauge field couples is an auxiliary field decoupled from the inflaton, some degree of nongaussianities leak into the curvature perturbations. This effect is however reduced with respect to the case where the inflaton is directly coupled to the photons by a factor $\epsilon\,N_\sigma$, with $\epsilon$ the slow roll parameter and $N_\sigma$ the number of e-folds while the auxiliary field is rolling. In particular, a factor $(\epsilon\,N_\sigma)^2$ enters in the curvature spectrum, and a factor $(\epsilon\,N_\sigma)^3$ enters in the curvature bispectrum.

In~\cite{Caprini:2014mja} two of us proposed a model able to provide magnetic seeds satisfying the lower bounds on the magnetic field intensity by gamma-ray observations and able to start the dynamo leading to the observed galactic fields while keeping in the perturbative regime. The model is based on the combination of two couplings: the Lagrangian presents both a term $I^2(\sigma(\tau))\,F_{\mu\nu}F^{\mu\nu}/4$ and a term $I^2(\sigma(\tau))\,\gamma\, \eta_{\mu\nu\rho\lambda}F^{\mu\nu}F^{\rho\lambda} /8 $, where $I(\sigma(\tau))=(-H\tau)^n$, $H$ is the physical Hubble scale, $\tau$ denotes conformal time, and $\gamma$ is a dimensionless constant of order $10$ or so. The function $I(\sigma(\tau))$ insures the coupling with the expanding background, and $\sigma(\tau)$ can be the inflaton or an auxiliary field. We restrict to $-2<n<0$: $n<0$ insures that the model is healthy from the point of view of strong coupling, while $n>-2$ that it is healthy from the point of view of backreaction. Note that this last condition is sufficient if the scalar field $\sigma(\tau)$ is the inflaton, but not if it is an auxiliary field, which is the case of interest for us. At the end of section \ref{sec:constraints} we analyse the (loose) conditions under which backreaction on the dynamics of $\sigma(\tau)$ is avoided in our scenario.

The model exploits the advantages of both couplings: the result is a helical magnetic field with large amplitude that can undergo inverse cascade and be amplified at large scales during the radiation dominated era~\cite{Banerjee:2004df,Durrer:2013pga} (thanks to the second term), but with a spectrum which is less steep than in the case of the pure axion-like coupling, therefore allowing interesting amplitudes at large scales (thanks to the first term) \cite{Durrer:2010mq}. To be more specific, while the axial coupling allows to generate fields that satisfy the blazar constraint, it is the coupling to $I^2(\sigma(\tau))\,F_{\mu\nu}F^{\mu\nu}/4$ that guarantees that the produced field can provide the required seeds for the observed galactic fields. In fact, the dynamics of the inverse cascade establishes a relationship between the root mean square amplitude (i.e. the intensity) of the field and its correlation length today (see eqs.~\eqref{eqs:inverse_cascade} below), so that, once an intensity of the field that satisfies the blazar constraint is determined, then also its correlation length is fixed. In our model, as typical in the case of primordial generation mechanisms, such a correlation length is small: of the order of the parsec or so, much shorter than the typical $\mathcal{O}(\rm Mpc)$ correlation scale required to initiate the galactic dynamo. Furthermore, a purely axial coupling would yield a field with a blue spectral index~\cite{Anber:2006xt}, so that fields strong enough to satisfy the blazar constraints would be too weak at sufficiently large scales for the dynamo. The possibility of tuning the spectral index to redder values, provided in our model by the $I^2(\sigma(\tau))\,F_{\mu\nu}F^{\mu\nu}/4$ term, can make the field strong enough also at the Mpc scale.

As the work~\cite{Caprini:2014mja} has been completed before~\cite{Ferreira:2014zia}, it was assumed that the model was safe from the point of view of CMB constraints on non-gaussianity (see~\cite{Ballardini:2014jta} for a compendium of the effects an helical magnetic field can have on the CMB). In~\cite{Caprini:2014mja} we therefore only evaluated the tensor spectrum produced in our model: the interest of it being that the gauge field could give rise to sizable tensor modes even if inflation occurs at a low energy scale, since the model evades the Lyth bound. 

Here we therefore complete the analysis of \cite{Caprini:2014mja} in light of \cite{Ferreira:2014zia}. After a brief overview of the process of gauge field amplification in section \ref{sec:gauge}, we rederive the results of \cite{Ferreira:2014zia} (see also~\cite{Namba:2015gja}) in terms of the equation of motion for the inflation perturbation in flat gauge (section \ref{sec:equation:motion}). We then demonstrate in section \ref{sec:curvature} that the comoving curvature perturbation $\mathcal{R}$ at the end of inflation can indeed be related to the inflation perturbation in flat gauge via $\mathcal{R} \simeq (\mathcal{H}/\varphi_0')\delta \varphi_{\rm flat}$. This allows us to evaluate the curvature spectrum (section \ref{sec:spectrum}) and bispectrum (section \ref{sec:bispectrum}). We impose CMB constraints on the resulting $f_{\rm NL}^{\rm equil}$ and find that the model is indeed able to provide strong enough magnetic seeds to satisfy the bounds by gamma-ray observations \cite{Durrer:2013pga} without overproducing non-gaussianity in the primordial scalar perturbations (section \ref{sec:constraints}). However, for this to be the case, some conditions must hold: the energy scale of inflation depends on the parameter $n$, on the number of efolds of the auxiliary field $N_\sigma$, and on the number of efolds between the time the field stops rolling and the end of inflation $\Delta N$. We find that, for $N_\sigma\gtrsim N_{\rm obs}$ (where $N_{\rm obs}$ denotes the observable number of efolds) and for $\Delta N\lesssim  1$, the energy scale of inflation is in the range $10^6 
\lesssim \rho_{\rm inf}^{1/4} \lesssim 10^8$ GeV. Furthermore, if the spectral index is red enough, we find that the magnetic seed can also initiate the galactic dynamo.

The metric is the FLRW one in conformal time $ds^2=a(\tau)^2\,(-d\tau^2+d{\bf x}^2)$. A prime denotes derivative with respect to conformal time, while a dot denotes derivatives with respect to physical time. Greek indexes run from $0$ to $3$ while Latin ones are spatial. The totally antisymmetric tensor $\eta_{\mu\nu\rho\lambda}$ is such that $\eta_{0123}=-\sqrt{-g}$. $M_{pl}$ denotes the reduced Planck mass. During inflation, $a(\tau)\simeq 1/(-H\tau)^{1+\epsilon_\varphi+\epsilon_\sigma}$ (where we have assumed $\epsilon_\phi+\epsilon_\sigma\ll 1$) with the comoving Hubble parameter $\mathcal{H}\simeq-(1+\epsilon_\varphi+\epsilon_\sigma)/\tau$, and $\epsilon_\varphi+\epsilon_\sigma= (\mathcal{H}^2-\mathcal{H}')/\mathcal{H}^2$.


\section{The model}%

We consider a model described by the following Lagrangian~\cite{Caprini:2014mja}:
\begin{equation}\label{lagrangian}
	L=-\frac{1}{2}\nabla_{\mu}\varphi\nabla^{\mu}\varphi-V\left(\varphi\right)-\frac{1}{2}\nabla_{\mu}\sigma\nabla^{\mu}\sigma-U\left(\sigma\right)+I^{2}\left(\sigma\right)\left(-\frac{1}{4}F_{\mu\nu}F^{\mu\nu}+\frac{\gamma}{8}\eta_{\mu\nu\rho\lambda}F^{\mu\nu}F^{\rho\lambda}\right)
\end{equation}
where $\varphi$ is the inflaton, $\sigma$ is an auxiliary field, $F_{\mu\nu}=\nabla_{\mu}A_{\nu}-\nabla_{\nu}A_{\mu}$, $\eta_{\mu\nu\rho\lambda}$ is the totally antisymmetric tensor, and $V$ and $U$ are the potential of the inflaton and auxiliary field respectively. The field $\varphi$ drives the background dynamics, while $\sigma$ will be assumed to roll at constant velocity until it stops and decays towards the end of inflation. As anticipated in the Introduction, we couple the gauge field to $\sigma$ and not directly to the inflaton in order for the model not to overproduce nongaussianities, see section \ref{sec:one:field}.

We set $I\left(\sigma\left(\tau\right)\right)=\left(-H \tau\right)^{-n}$, where $H$ the physical Hubble parameter and $n<0$ a free parameter of the model, together with the dimensionless constant $\gamma$. The field $\sigma$ and the gauge field $A_{\mu}$ are minimally coupled to the inflaton. We consider $A_{\mu}$ as a quantity of the order of $\mathcal{O}\left(1/2\right)$ in cosmological perturbation theory, so that it does not contribute to the background dynamics. Quantities that are quadratic in the gauge field, such as the electromagnetic energy density, are therefore first order as the inflaton perturbations. 

The dynamics of the gauge field is non-standard. Because of the presence of the term $-\frac{I^{2}\left(\sigma\right)}{4}F_{\mu\nu}F^{\mu\nu}$, the spectral index of the magnetic field amplified by the rolling of $\sigma$ depends on the parameter $n$ and can in principle be red. On the other hand, the term $\frac{I^{2}\left(\sigma\right)}{8}\,\gamma\,\eta_{\mu\nu\rho\lambda}F^{\mu\nu}F^{\rho\lambda}$ induces an exponential amplification of one of the two helicities of the gauge field. This has the double effect of increasing the overall amplitude of the gauge field and of guaranteeing that only one of its two helicity modes is amplified, leading to a helical magnetic field. Helical magnetic fields undergo the inverse-cascade process \cite{Son:1998my,Field:1998hi,Vachaspati:2001nb,Sigl:2002kt,Christensson:2000sp,Campanelli:2007tc,Campanelli:2013iaa,Banerjee:2004df}, which further enhances the magnetic field amplitude at large scales during the subsequent epochs of the evolution of the universe. 

The above properties: a tunable spectral index, a tunable amplitude and amplification during radiation domination, guarantee that the model works in the context of a controllable theory that always remains weakly coupled. As explained in \cite{Caprini:2014mja}, in order to avoid strong coupling during inflation~\cite{Demozzi:2009fu}, the parameter $n$ is bounded to $n<0$. On the other hand, in order to avoid backreaction of the infrared modes of the gauge field on the background dynamics \cite{Martin:2007ue} (see also~\cite{Bartolo:2012sd,Kanno:2009ei}), one must impose $n>-2$. In summary then we restrict $-2<n<0$ ($n=0$ is excluded since in such a case there is no magnetic field production). At the end of section \ref{sec:constraints} we analyse other conditions under which the absence of backreaction is guaranteed in the present scenario.

We define the parameter $\xi\equiv -n\,\gamma$ that, as we will see, quantifies the enhancement of the magnetic field amplitude. 
In the following we review the dynamics of $A_\mu$.

\subsection{Amplification of helical magnetic fields}
\label{sec:gauge}

Here we summarize the results about the magnetic field production for the model \eqref{lagrangian}; see \cite{Caprini:2014mja} for more details.

We define the canonically normalized $\tilde{A}_\mu=I\,A_\mu$, that we quantize, in the Coulomb gauge $\tilde{A}_0=\partial_i\,\tilde{A}_i=0$, as usual:
\begin{equation}\label{eq:quant_A}
	\tilde{\textbf{A}}_{i}\left(\textbf{x}\right)=\sum_{\lambda=\pm}\int\frac{d^{3}\textbf{k}}{\left(2\pi\right)^{3/2}}\varepsilon_{i}^{\lambda}\left(\textbf{k}\right)e^{i\textbf{k}\cdot \textbf{x}}\left[\tilde{A}_{\lambda}\left(\textbf{k},\tau\right)\hat{a}_{\lambda}\left(\textbf{k}\right)+\tilde{A}_{\lambda}^{\ast}\left(-\textbf{k},\tau\right)\hat{a}_{\lambda}^{\dag}\left(-\textbf{k}\right)\right]\,,
\end{equation}
where $\varepsilon_{i}^{\lambda}\left(\textbf{k}\right)$ is the helicity vector, and $\lambda=\pm$ indicates the helicity state. Varying the Lagrangian \eqref{lagrangian}, the equation of motion of $\tilde{A}_{\lambda}$ results:
\begin{equation}\label{eqmoto:a}
	\tilde{A}_{\lambda}''+\left[-\frac{n\left(n+1\right)}{\tau^{2}}+2\,\lambda\,\xi\,\frac{k}{\tau}+k^{2}\right]\,\tilde{A}_{\lambda}=0\,,
\end{equation}
where the prime indicates derivative with respect to conformal time. 

The system evolves in three stages. For $\tau\rightarrow -\infty$ the term $k^2$ dominates and the photons are in the Bunch-Davies vacuum. Somewhat before horizon crossing $|k\,\tau|\lesssim \xi$, when particle production starts occurring, the sign of the second term in the parenthesis matters: it depends on the helicity $\lambda$, and since $\tau<0$ only the $\lambda=+$ helicity mode is amplified. At this stage a net chirality in the gauge field is generated. Finally, as $\tau\rightarrow 0^-$, the first term in the parenthesis takes over. The final result of this process is a field with a spectral index controlled by the parameter $n$ and with net helicity. 
In order for an efficient enhancement to take place, $\xi\gg 1$ is required.

For $\left|k\,\tau\right|\ll\xi$, that is for the regime in which the modes are outside of the Bunch-Davies vacuum, the solution of eq.~\eqref{eqmoto:a} reads \cite{Durrer:2010mq,Caprini:2014mja}
\begin{equation}\label{sol:a:full}
	\tilde{A}_{+}\left(k,\tau\right)\simeq\sqrt{-\frac{2\,\tau}{\pi}}e^{\pi\,\xi}K_{-2n-1}\left(\sqrt{-8\,\xi\, k\, \tau}\right)\,,
\end{equation}
where $K_{\alpha}$ is the modified Bessel function ($\tilde{A}_+\left(\textbf{k},\tau\right)=\tilde{A}_+\left(k,\tau\right)$, so from now on we indicate only the dependence on the norm of $\textbf{k}$).
For $|k\,\tau| \ll 1/\xi$ the above expression reduces to
\begin{equation}\label{sol:a:approx}
	\tilde{A}_{+}\left(k,\tau\right)\simeq\sqrt{-\frac{\tau}{2\,\pi}}\,e^{\pi\,\xi}\,\Gamma\left(\left|2n+1\right|\right)\left|2\,\xi\, k\, \tau\right|^{-\left|n+1/2\right|}\,.
\end{equation}
The amplification of the gauge field is therefore exponential in the parameter $\xi$. 

We define the electric and magnetic field in real space as 
\begin{align}\label{eq:def_eb}
E_i({\bf x})\equiv -\frac{A_i'({\bf x})}{a^{2}}\,,\qquad\qquad B_i({\bf x})=\epsilon_{ijl} \,\frac{\partial_j A_l({\bf x})}{a^{2}}\,,
\end{align}
where $\epsilon_{ijl}$ is the totally antisymmetric tensor in flat space\footnote{Note that for convenience we insert in these definitions an extra factor $a^
{-1}$ with respect to the standard definitions arising from the field tensor, that would read: $E_\mu=u^\nu F_{\mu\nu}=-\partial_0 A_\mu /a$ and $B_\mu=\eta_{\mu\nu\alpha\beta} \,u^\beta F^{\nu\alpha}/2=\eta_{\mu\nu\alpha\beta} \,u^\beta (\partial^\nu A^\alpha - \partial^\alpha A^\nu)/2$.}. Defining the magnetic field power spectrum as usual as 
\begin{align}
\langle B_i({\bf k})\,B_j^*({\bf q})\rangle = \frac{1}{2}\, (2\pi)^3\, (\delta_{ij}-\hat{k}_i\hat{k}_j)\,P_B(k)\,\frac{\delta({\bf k}-{\bf q})}{k^3}\,,
\end{align}
with $\sqrt{P_B(k)}\propto k^{n_B}$, from Eq.~\eqref{sol:a:approx} we see that the magnetic spectral index at large scales is 
\begin{equation}\label{eq:nB}
	n_B=\frac{5}{2}-\left|n+\frac{1}{2}\right|\,.
\end{equation} 
Furthermore, defining the intensity of the magnetic field as
\begin{equation}
	B^{2}\equiv \langle B^{2}\rangle=\int\frac{d^{3}\textbf{k}}{\left(2\pi\right)^{3}}\left|k \,A_{+}\right|^{2}\,,
\end{equation}
the intensity of the magnetic field at the end of inflation is found to be \cite{Caprini:2014mja}
\begin{equation}
	B^{2}_{reh}=H^{4}\,\frac{e^{2\pi\,\xi}}{\xi^{5}}\frac{\Gamma\left(4-2n\right)\,\Gamma\left(6+2n\right)}{2^{8}\times 315\, \pi^{3}}\, e^{-4\,\Delta N}\,,
\end{equation}
where we have added, with respect to the result of \cite{Caprini:2014mja}, a factor $e^{-4\,\Delta N}$ that accounts for the redshifting of the magnetic field during inflation if the field $\sigma$ stops rolling $\Delta N$ efoldings before the end of inflation.

In \cite{Caprini:2014mja}, in particular in Figures 3 and 4, it was shown that the associated production of magnetic fields can satisfy the lower bounds reported in \cite{Neronov:1900zz,Taylor:2011bn,Vovk:2011aa}, keeping under control the related gravitational wave production, and without accounting for the $e^{-4\,\Delta N}$ factor included above. Moreover, in \cite{Caprini:2014mja} it resulted that $\xi$ is of the order of $\mathcal{O}\left(10\right)$ for the interesting parameter region. 

Our analysis relies on the validity of perturbation theory. As pointed out in~\cite{Ferreira:2015omg}, one should make sure that higher order contributions to the two-point function of the gauge field are much smaller than the leading term extracted from eq.~(\ref{sol:a:full}). As discussed in~\cite{Peloso:2016gqs} for what concerns the related system where only the operator $F_{\mu\nu}\,\tilde{F}^{\mu\nu}$ is coupled to the inflaton, the validity of perturbation theory is essentially guaranteed when the energy in the modes of the gauge field is much smaller than the energy in the rolling scalar (the kinetic energy $\simeq \dot\sigma^2/2$, in our case). In Section 6.1 below we will discuss the region of parameter space where such a condition is satisfied in our system. We believe that, as long as we confine ourselves to that portion of parameter space, the perturbative analysis leading to eq.~(\ref{sol:a:full}) can be trusted.

Before concluding this Section let us note that despite the large occupation number attained by the gauge field, we do not expect photon scatterings~\cite{Ferreira:2017lnd} to play any important role because of the smallness of the gauge coupling $\sim I^{-1}\propto e^{-|n|\,N}$, where $N$ is the number of efoldings before the end of inflation.

\section{Perturbations of the scalar fields}

The gravitational coupling of the inflaton to the auxiliary field and the gauge field gives rise to active sources in the equation of motion of the inflaton perturbations. Therefore we expect an extra contribution to the curvature perturbations besides those due to vacuum fluctuations of the inflaton. Moreover, the gauge field also induces a source term in the equation of motion of the tensor modes, leading to an additional population of gravitational waves on the top of those generated by quantum fluctuations of the gravitational field. The extra production of scalar and tensor modes depends on the parameters of model~\eqref{lagrangian}.
Therefore we expect current observational constraints on scalar and tensor perturbations to provide bounds on the parameters of our model~\cite{Barnaby:2010vf,Barnaby:2012xt,Mukohyama:2014gba,Namba:2015gja}. \\

\subsection{Tensor modes}
\label{sec:tensor}

The gauge field is an additional source of gravitational waves~\cite{Sorbo:2011rz,Barnaby:2012xt,Mukohyama:2014gba,Namba:2015gja}, besides the vacuum ones, that violate the Lyth bound. Since efficient magnetogenesis typically implies a low energy scale inflation, the gravitational wave signal actively sourced by the gauge field is in general more significant than that due to quantum fluctuations of the gravitational field. The bounds on the parameters of the present model due to the extra production of tensor modes have been evaluated in \cite{Caprini:2014mja}. In particular, imposing that the magnetic field intensity corresponds to the lower bound from gamma-ray observations allows one to reduce the analysis to one free parameter $n$, in terms of which one derives the allowed energy scale of inflation. In \cite{Caprini:2014mja} a tensor-to-scalar ratio entirely due to the gauge field of $r=0.2$ was assumed. This is now excluded by current constraints~\cite{Array:2015xqh}. Ref.~\cite{Caprini:2014mja} showed that such a value of $r$ would constrain the energy scale of inflation to $10^5 < \rho_{\rm inf}^{1/4} < 5\cdot 10^{10}$ GeV, as a function of $-2<n<0$ (c.f. Fig.~3 of \cite{Caprini:2014mja}). At the end of this work we re-evaluate $r$ in light of our new results: c.f. section \ref{sec:constraints}. 

\subsection{Scalar modes}

In this Section we compute the contribution of the gauge field to curvature perturbations. We expect them to provide stricter limits on the parameter space of the model with respect to those due to tensor perturbations: i.e., we expect the allowed energy scale of inflation to be lower. However, as we will see, constraints from the curvature power spectrum and from the bounds on nongaussianities, in combination with successful magnetogenesis, still allow inflation to happen at reasonable energy scales.

\subsubsection{Equation of motion of the inflaton perturbations}
\label{sec:equation:motion}

We perturb the inflaton and the auxiliary field as usual, $\varphi=\varphi_0+\delta\varphi$ and $\sigma=\sigma_0+\delta\sigma$, where $\delta\varphi$ and $\delta\sigma$ are first order quantities while $A_\mu$ is half order one. Here we identify the equation of motion of $\delta\varphi$ which is then required in order to evaluate the curvature perturbations.

We work in the flat gauge and follow the conventions of \cite{Malik:2008im}. From the Lagrangian~\eqref{lagrangian} we use Einstein equations to trade the metric perturbations for the matter perturbations~\cite{Barnaby:2012xt} and we obtain the equation for $\delta\varphi$ by subtracting from the exact equation for $\varphi$ the equation for the zero mode $\varphi_0$, where the perturbation in the gauge field is evaluated in Hartree approximation. Using this procedure, the equation of motion of $\delta\varphi$ in momentum space reads
\begin{align}\label{eqmoto:phi}
	\delta\varphi_{\rm flat}''+2\,\mathcal{H}\,\delta\varphi_{\rm flat}'+\left(k^{2}+a^{2}\,V_{\varphi\varphi}\right)\,\delta\varphi_{\rm flat}-&\left(\frac{a^{2}\,\varphi'^{2}}{\mathcal{H}}\right)'\,\frac{\delta\varphi_{\rm flat}}{M_{pl}^{2}a^{2}}-\left(\frac{a^{2}\,\varphi'\,\sigma'}{\mathcal{H}}\right)'\frac{\delta\sigma_{\rm flat}}{M_{pl}^{2}\,a^{2}}=\nonumber \\
	&=2\,\varphi'_{0}\,S^{\left(3\right)}+\frac{\varphi'_{0}}{\mathcal{H}}\,S'^{\left(3\right)}+\frac{\varphi'_{0}}{\mathcal{H}}\,S^{\left(2\right)}
\end{align}
where $V_{\varphi\varphi}$ indicates the second derivative of the potential with respect to $\varphi$, and $\mathcal{H}$ is the comoving Hubble rate. The source terms depend on the gauge field and can be written in terms of the electromagnetic energy density and Poynting vector as follows
\begin{eqnarray}
	S^{\left(2\right)}&=&-\frac{a^2}{2\,M_{pl}^2}\,\rho_{\rm em}({\bf k})= -\frac{I^{2}\,a^2}{4\,M_{pl}^{2}}\,[E_i*E_i+B_i*B_i] \label{eq:S2} \\
	S^{\left(3\right)}&=& \frac{a}{2\,M_{pl}^2} \frac{i\,\hat{k}_j}{k}\, q_{{\rm em}j}({\bf k})= \frac{I^2\,a^2}{2\,M_{pl}^2} \frac{i\,\hat{k}_j}{k} \,\epsilon_{jlm}[E_l * B_m]\,,  \label{eq:S3}
\end{eqnarray}
where the convolution is defined as
\begin{equation}
	[E_l * B_m]({\bf k}) = \int \frac{d^3q}{(2\pi)^{3/2}} \,E_l({\bf k-q}) B_m({\bf q})\,.
\end{equation}
Note that a term involving $\delta\sigma_{\rm flat}$ is present in the equation of motion of $\delta\varphi_{\rm flat}$. This term and the sources of the r.h.s of eq.~\eqref{eqmoto:phi} come from the metric perturbations. Without accounting for these terms, one would find that $\delta\varphi_{\rm flat}$ satisfies the Klein-Gordon equation, since the inflaton $\varphi$ is only minimally coupled to the other fields.
Analogously, the equation of motion for $\delta\sigma_{\rm flat}$ reads
\begin{align}\label{eqmoto:sigma}
	\delta\sigma_{\rm flat}''+2\,\mathcal{H}\,\delta\sigma_{\rm flat}'+\left(k^{2}+a^{2}\,V_{\sigma\sigma}\right)\,\delta\sigma_{\rm flat}-&\left(\frac{a^{2}\sigma'^{2}}{\mathcal{H}}\right)'\frac{\delta\sigma_{\rm flat}}{M_{pl}^{2}\,a^{2}}-\left(\frac{a^{2}\,\varphi'\,\sigma'}{\mathcal{H}}\right)'\frac{\delta\varphi_{\rm flat}}{M_{pl}^{2}\,a^{2}}= \nonumber\\
	&=S^{\left(1\right)}+2\,\sigma'_{0}\,S^{\left(3\right)}+\frac{\sigma'_{0}}{\mathcal{H}}\,S'^{\left(3\right)}+\frac{\sigma'_{0}}{\mathcal{H}}\,S^{\left(2\right)}\,,
\end{align}
where
\begin{equation}\label{eq:S1}
	S^{\left(1\right)}=a^{2}II_{\sigma}\left[E_i*E_i-B_i*B_i+2\,\gamma\,\left(E_i*B_i\right)\right]\,,
\end{equation}
and $I_{\sigma}\equiv {\rm d}I/{\rm d}\sigma$. 

Equations \eqref{eqmoto:phi} and \eqref{eqmoto:sigma} form a coupled system due to the mixing terms, that we now proceed to solve by diagonalising it. We account for both the mixing terms and the sources originated by the gravitational coupling between the inflaton and the other fields, i.e. we solve the exact equations at leading order in slow-roll. For compactness we denote $S_{\varphi}$ the source on the r.h.s of eq.~\eqref{eqmoto:phi}, and $S_{\sigma}$ the source on the r.h.s. of eq.~\eqref{eqmoto:sigma}.
We express the system \eqref{eqmoto:phi}-\eqref{eqmoto:sigma} with respect to $u_{\varphi}\equiv a\,\delta\varphi$ and $u_{\sigma}\equiv a\,\delta\sigma$. At first order in slow-roll we have 
\begin{align}\label{system:u}
&u_{\varphi}''+\left(k^{2}-\frac{2}{\tau^{2}}\right)\,u_{\varphi}-\frac{3}{\tau^{2}}\left(3\,\epsilon_{\varphi}-\eta_{\varphi}\right)\,u_{\varphi}-\frac{3}{\tau^{2}}\left(2\,\Theta\,\epsilon_{\varphi}\right)\,u_{\sigma}=a\left(\tau\right)\,S_{\varphi}\,,\\
&	u_{\sigma}''+\left(k^{2}-\frac{2}{\tau^{2}}\right)\,u_{\sigma}-\frac{3}{\tau^{2}}\left(\epsilon_{\varphi}+2\,\Theta^2\,\epsilon_{\varphi}-\eta_{\sigma}\right)\,u_{\sigma}-\frac{3}{\tau^{2}}\left(2\,\Theta\,\epsilon_{\varphi}\right)\,u_{\varphi}=a\left(\tau\right)\,S_{\sigma}\,,
\end{align}
where we have defined $\epsilon_\varphi=\dot\varphi^2/(2\,H^2\,M_{pl}^2)$, $\eta_{\varphi}\equiv V_{\varphi\varphi}/V$ (and analogously for $\eta_{\sigma}$), and $\Theta\equiv\sigma_{0}'/\varphi_{0}'$, so that $\dot{H}/H^2=-\epsilon_\varphi\left(1+\Theta^2\right)$ (we remind the reader that we use a prime for derivatives w.r.t. conformal time, and a dot for derivatives w.r.t. physical time). As we discuss below, we will consider the regime $\epsilon_{\varphi}\ll \Theta^2\,\epsilon_{\varphi}\ll 1$, where the latter inequality is needed to ensure that we are in an inflating regime, $|\dot{H}|\ll H^2$. The mixing of $u_\varphi$ with $u_\sigma$ is described by the following matrix
\begin{equation}
M_{\varphi\sigma}=\label{eq:M}
\begin{pmatrix}
	2+9\,\epsilon_{\varphi}-3\,\eta_{\varphi} & 6\,\Theta\,\epsilon_{\varphi}\\
	6\,\Theta\,\epsilon_{\varphi} & 2+3\,\epsilon_{\varphi}\left(1+2\,\Theta^2\right)-3\,\eta_{\sigma}
\end{pmatrix}
\end{equation}
which can be diagonalized by $U\cdot M\cdot U^{T}=\Lambda={\mathrm {diag}}\left(\lambda_{\varphi},\,\lambda_{\sigma}\right)$, with $U\left(\theta\right)$ a rotation matrix. We define $\mathbf{v}\equiv U^{-1}\mathbf{u}$, i.e. the eigenvectors of the $\Lambda$ matrix. The diagonalized system then reads
\begin{equation}\label{system:rotated:varphi}
	v_{\varphi}''+\left(k^{2}-\frac{\lambda_{\varphi}}{\tau^{2}}\right)\,v_{\varphi}=a(\tau)\,[\cos\theta\, S_{\varphi}+\sin\theta \,S_{\sigma}]\,,
\end{equation}
\begin{equation}\label{system:rotated:sigma}
	v_{\sigma}''+\left(k^{2}-\frac{\lambda_{\sigma}}{\tau^{2}}\right)v_{\sigma}=a(\tau)\,[-\sin\theta\, S_{\varphi}+\cos\theta\, S_{\sigma}]\,. 
	\end{equation}
The solutions for $v_{\varphi}$ and $v_{\sigma}$ are then given by
\begin{align}\label{v}
	&v_{\varphi}\left(\textbf{k},\tau\right)=\int d\tau'\, G_{k}^{\lambda_{\varphi}}\left(\tau,\tau'\right)\,a(\tau')\,\left[\cos\theta \,S_{\varphi}+\sin\theta\, S_{\sigma}\right]\,,\\
	&v_{\sigma}\left(\textbf{k},\tau\right)=\int d\tau' \,G_{k}^{\lambda_{\sigma}}\left(\tau,\tau'\right)\,a(\tau')\,\left[-\sin\theta \,S_{\varphi}+\cos\theta\, S_{\sigma}\right]\,,
\end{align}
where $G_{k}^{\lambda_{\sigma/\varphi}}$ are the Green functions of the equations \eqref{system:rotated:varphi} and \eqref{system:rotated:sigma} (note that the dependence on ${\bf k}$ and $\tau'$ is understood in the sources $S_{\varphi}$, $S_{\sigma}$). From the definition of $v_{\varphi}$, we can go back to the original function
\begin{align}\label{u:green:sorgenti}
	u_{\varphi}\left(\textbf{k},\,\tau\right)=& \,\,v_{\varphi}\,\cos\theta -v_{\sigma}\,\sin\theta= \nonumber\\
	=&\int d\tau'\, a(\tau')\,\left\{G_{k}^{\lambda_{\varphi}}\left(\tau,\,\tau'\right)\,\left[\cos^2\theta\,S_{\varphi}+\sin\theta\,\cos\theta\, S_{\sigma}\right]+\nonumber \right.\\
	& \left. +G_{k}^{\lambda_{\sigma}}\left(\tau,\,\tau'\right)\left[\sin^2\theta\,S_{\varphi}-\sin\theta\,\cos\theta \,S_{\sigma}\right]\right\}\,.
\end{align}
Note that for our purposes it is enough to evaluate the variable $u_{\varphi}\left(\textbf{k},\tau\right)$, that is connected to $\delta\varphi$, since ultimately we are interested in calculating the curvature perturbation (c.f. section \ref{sec:curvature}). 
The Green function is given by
\begin{eqnarray}\label{green:full}
G_{k}^{\lambda_{\varphi/\sigma}}\left(\tau,\tau'\right)=\frac{\pi}{2}\sqrt{\tau\tau'}\left[J_{\mu}\left(-k\tau\right)\,Y_{\mu}\left(-k\tau'\right)-J_{\mu}\left(-k\tau'\right)\,Y_{\mu}\left(-k\tau\right)\right]\,\Theta\left(\tau-\tau'\right) \\
\quad \mbox{with} \quad \mu=\frac{1}{2}\sqrt{1+4\,\lambda_{\varphi/\sigma}}\,,\nonumber
\end{eqnarray}
where $\Theta$ here denotes the Heaviside step function. The relevant solution for the field perturbation concerns modes outside the horizon, so that $\left|k\,\tau\right|\ll 1$. Furthermore, the solution for the gauge field \eqref{sol:a:full} entering in the sources $S_\varphi$ and $S_\sigma$ is exponentially suppressed for $\left|k\,\tau'\right|\gtrsim 1/\xi\ll 1$, so that  the biggest contribution to the above integral comes from modes well outside the horizon and we can set $\left|k\,\tau'\right|\ll 1$ in the Green function. Therefore, we can approximate the Green function in \eqref{green:full} as
\begin{equation}\label{eq:Green_over_horizon}
	G^{\lambda_{\varphi/\sigma}}_{k}\left(\tau,\,\tau'\right)\simeq \frac{\sqrt{\tau\,\tau'}}{2\,\mu}\left(\frac{\tau'}{\tau}\right)^{\mu}\,.
\end{equation}
We next exploit the slow-roll expansion. The system matrix given in~\eqref{eq:M} has the form
\begin{equation}
M_{\varphi\sigma}=
	\begin{pmatrix}
	2+\delta M_{11} & \delta M_{12}\\
	\delta M_{12} & 2+\delta M_{22}
	\end{pmatrix}\,,
\end{equation}
where $\left|\delta M_{\varphi\sigma}\right|$ are quantities that are first-order in slow-roll. The eigenvalues of $M_{\varphi\sigma}$ are
\begin{equation}
	\lambda_{\varphi/\sigma}=2+\frac{1}{2}\left[\delta M_{11}+\delta M_{22}\pm \sqrt{4\,\delta M_{12}^{2}+\left(\delta M_{11}-\delta M_{22}\right)^{2}}\right]\,, 
\end{equation}
so that we can define the quantities $\delta\lambda_{\sigma/\varphi}$, also first order in slow roll, by
\begin{equation}
	\Lambda=
	\begin{pmatrix}
	\lambda_{\varphi} & 0\\
	0 & \lambda_{\sigma}
	\end{pmatrix}
	=
	\begin{pmatrix}
	2+\delta\lambda_{\varphi} & 0\\
	0 & 2+\delta\lambda_{\sigma}
	\end{pmatrix}
	\,.
\end{equation}
As a consequence one can also expand the Bessel function index of eq.~\eqref{green:full} as $\mu\simeq 3/2+\delta\lambda_{\varphi/\sigma}/3$, which, applied to eq.~\eqref{eq:Green_over_horizon}, leads to the following expansion for the Green function:
\begin{equation}
	G_{k}^{\lambda_{\varphi/\sigma}}\simeq G^{(2)}_{k}+\delta\lambda_{\varphi/\sigma}\,\tilde{G}^{(2)}_{k}\equiv \frac{\sqrt{\tau\,\tau'}}{3}\left(\frac{\tau'}{\tau}\right)^{3/2}\left(1-\frac{\delta\lambda_{\varphi/\sigma}}{3}+\frac{\delta\lambda_{\varphi/\sigma}}{3}\,{\rm log}\frac{\tau'}{\tau}\right) \,.
\end{equation}
The term proportional to the logarithm gives the dominant contribution between the last two terms, therefore we can use
\begin{equation}\label{eq:Green}
	G_{k}^{(2)}\left(\tau,\tau'\right)=\frac{\sqrt{\tau\,\tau'}}{3}\left(\frac{\tau'}{\tau}\right)^{3/2}\,,\qquad \tilde{G}_{k}^{(2)}\simeq \frac{\sqrt{\tau\,\tau'}}{9}\left(\frac{\tau'}{\tau}\right)^{3/2}\,\log\left(\frac{\tau'}{\tau}\right)\,,
\end{equation}
which, applied to \eqref{u:green:sorgenti} finally leads to
\begin{eqnarray}
	u_{\varphi}\left({\bf k},\tau\right)&\simeq& \sin\theta\,\cos\theta\,\left(\delta\lambda_{\varphi}-\delta\lambda_{\sigma}\right)\,\int d\tau' \,a(\tau')\,\tilde{G}_{k}^{\left(2\right)}\left(\tau,\,\tau'\right)\,S_{\sigma}\\
	&&+\int d\tau' \,a(\tau') \,G_{k}^{\left(2\right)}\left(\tau,\tau'\right)\,S_{\varphi}\,.
\end{eqnarray}
Note that, in the last term, we have neglected a term of order $\delta\lambda_{\varphi/\sigma}\,S_{\varphi}$, since, as we will see below, $S_\varphi$ is higher order in slow roll with respect to $S_\sigma$.

Now we can make a further simplification, which consists in approximating the logarithmic term ${\rm log}\left({\tau'}/{\tau}\right)$ appearing in $\tilde{G}_{k}^{\left(2\right)}$ simply by a factor $N_{\sigma}$ that we multiply outside of the time integral, where $N_{\sigma}$ is the number of e-folds during which the field $\sigma$ is rolling. This is possible because the integral in $d\tau'$ is dominated by $|\tau'|\sim (k\,\xi)^{-1}$, and greatly simplifies things, since the remaining part of the Green function actually corresponds to the one of eqs.~\eqref{system:u} at zeroth order in slow roll. In summary, we can reduce the problem of determining the inflaton perturbation in the model under analysis to solving the following equation of motion for $u_{\varphi}=a\,\delta\varphi$:
\begin{equation}\label{eqmoto:u:sorgenti}
	u_{\varphi}''+\left(k^{2}-\frac{2}{\tau^{2}}\right)\,u_{\varphi}=\frac{N_{\sigma}}{3}\,\sin\theta\,\cos\theta\, \left(\delta\lambda_{\varphi}-\delta\lambda_{\sigma}\right)a\left(\tau\right)S_{\sigma}+a\left(\tau\right)S_{\varphi}\,.
\end{equation}
This equation presents an effective source, combination of those for the fluctuations of the inflaton and of the auxiliary field (c.f. eqs.~\eqref{eqmoto:phi} and \eqref{eqmoto:sigma}), that we now proceed to estimate.  

\subsubsection{Estimation of the source}

We want to evaluate the relative contribution of the source terms on the r.h.s of eq.~\eqref{eqmoto:u:sorgenti}. The exact relation
\begin{equation}
	\sin\theta\, \cos\theta =\frac{\delta M_{12}}{\sqrt{4\,\delta M_{12}^{2}+\left(\delta M_{11}-\delta M_{22}\right)^{2}}}\,,
\end{equation}
implies that
\begin{equation}
	\sin\theta\, \cos\theta\, \left(\delta\lambda_{\varphi}-\delta\lambda_{\sigma}\right)=\delta M_{12}=6\,\Theta \,\epsilon_{\varphi}\,.
\end{equation}

Going back to the full expressions given in eqs.~\eqref{eqmoto:phi} and \eqref{eqmoto:sigma}, the source in eq.~\eqref{eqmoto:u:sorgenti}, barring a overall factor $a\left(\tau\right)$, reads
\begin{equation}
2\,N_{\sigma}\,\Theta\,\epsilon_\varphi \,\left[S^{\left(1\right)}+2\,\sigma'_{0}\,S^{\left(3\right)}+\frac{\sigma'_{0}}{\mathcal{H}}\,S'^{\left(3\right)}+\frac{\sigma'_{0}}{\mathcal{H}}\,S^{\left(2\right)}\right]+ 2\,\varphi'_{0}\,S^{\left(3\right)}+\frac{\varphi'_{0}}{\mathcal{H}}\,S'^{\left(3\right)}+\frac{\varphi'_{0}}{\mathcal{H}}\,S^{\left(2\right)}\,,
\end{equation}
that, collecting terms and considering that $\Theta\,\left|\frac{\sigma'_0}{\varphi_0'}\right|\,\epsilon_\varphi=\Theta^2\,\epsilon_\varphi\ll N_\sigma^{-1}\ll 1$, can be approximated to
\begin{align}\label{sorgente:tot}
2\,N_{\sigma}\,\Theta\,\epsilon_\varphi \,S^{\left(1\right)}+2\,\varphi_{0}'\,S^{\left(3\right)}+\frac{\varphi_{0}'}{\mathcal{H}}\,S'^{\left(3\right)}+\frac{\varphi_{0}'}{\mathcal{H}}\,S^{\left(2\right)}\,,
\end{align}
which amounts to neglecting the source due to the metric perturbations (as opposed to $S^{(1)}$, the one generated by the direct coupling between $\sigma$ and the gauge field) in the equation of motion of $\delta\sigma$, eq.~\eqref{eqmoto:sigma}.

Let us now compare the term proportional to $S^{\left(1\right)}$ to the other terms in eq.~\eqref{sorgente:tot}. To do this we first need to estimate the amplitude of the fields $E$ and $B$ on which the sources depend, c.f. eqs.~\eqref{eq:S2}, \eqref{eq:S3} and \eqref{eq:S1}. Let us go back to solution \eqref{sol:a:full} for the gauge field, applying it in the regime relevant for particle creation, i.e. for $\left|k\,\tau\right|\ll \xi$ and $\xi\gg 1$. We can use it to approximate 
\begin{align}
	&B(k, \tau)\sim k\, A\left(k,\,\tau\right)\simeq k \,\left(-H\tau\right)^{n}\sqrt{-\frac{2\tau}{\pi}}\,e^{\pi\,\xi}\,K_{-2n-1}\left(\sqrt{-8\,\xi \,k\,\tau}\right)\,,\\
	&E(k, \tau)\sim -A\left(k,\,\tau\right)'\simeq -\left(-H\,\tau\right)^{n}\sqrt{\frac{4\,\xi\, k}{\pi}}\,e^{\pi\,\xi}\,K_{-2n}\left(\sqrt{-8\,\xi\, k\,\tau}\right)\,.
\end{align}
Moreover, the largest contribution to the above expressions comes from modes with $-k\tau \sim 1/\xi$, since the Bessel functions are exponentially suppressed for values of their arguments that are much larger than unity and the phase space suppresses contributions with small $k$. In this regime the Bessel functions take values that are of the order of the unity and 
\begin{equation}\label{eq:E:B}
	\frac{E}{B}\sim \sqrt{\frac{\xi}{-k\,\tau}}\sim \xi\,,
\end{equation}
where in the last step we have used again $-k\,\tau\sim1/\xi$. This means that the magnetic field is suppressed by a factor $1/\xi$ with respect to $E$ (remember $\xi \gg 1$). Therefore we can conclude that in $S^{(1)}$ (eq.~\eqref{eq:S1}) the term proportional to the convolution $E_i*E_i$ is dominant with respect to the one proportional to $B_i*B_i$. 

Exploiting eq.~\eqref{eq:E:B}, we can parametrically estimate the amplitude of the terms in \eqref{sorgente:tot}. Whenever there is a residual dependence on the momentum, we use the value corresponding to the peak of the electric and magnetic fields, i.e. $|k\,\tau| \sim 1/\xi$. Using $I_{\sigma}\simeq n \,\mathcal{H}\,I/(\Theta\, \varphi_0')$, which is valid at leading order in $\epsilon_\varphi$ and $\Theta^2 \epsilon_\varphi$, one gets
\begin{align}
	& 2\,N_{\sigma}\,\Theta\, \epsilon_\varphi \, S^{\left(1\right)} \sim  2\,N_{\sigma}\,\Theta \,\epsilon_\varphi\, a^{2}\,n \,\mathcal{H}\,\frac{I^{2}}{\sigma_{0}'}(E*E)\simeq N_{\sigma}\,n\,\sqrt{2\epsilon_\varphi}\left(\frac{a^{2}I^{2}\,E*E}{M_{pl}}\right)\,,\nonumber\\
	&\varphi_{0}'\,S^{\left(3\right)}\sim \varphi_{0}'\,\frac{a^{2}\,I^{2}}{2\,M_{pl}^{2}}\,\frac{B*E}{k}\sim \sqrt{\frac{\epsilon_\varphi}{2}}\left(\frac{a^{2}I^{2}\,E*E}{M_{pl}}\right)\,,\nonumber\\
	&\frac{\varphi_{0}'}{\mathcal{H}}\,S'^{\left(3\right)}\sim \sqrt{\frac{\epsilon_\varphi}{2}}\,\left(\frac{a^{2}\,I^{2}\,E*E}{M_{pl}}\right)\,,\nonumber\\
	&\frac{\varphi_{0}'}{\mathcal{H}}\,S^{\left(2\right)}\sim \sqrt{\frac{\epsilon_\varphi}{8}}\left(\frac{a^{2}\,I^{2}\,E*E}{M_{pl}}\right)\,.
\end{align}

From the above estimations we conclude that all terms in eq.~\eqref{sorgente:tot} are of the same order of magnitude, therefore we cannot neglect any of them a priori. In summary, to evaluate the inflaton perturbation (and therefore the curvature power spectrum), one has to solve the following equation:
\begin{equation}\label{eqmoto:phi:approx}
	\delta\varphi''_{\rm flat}+2\,\mathcal{H}\,\delta\varphi'_{\rm flat}+k^{2}\,\delta\varphi_{\rm flat}= 2\,N_{\sigma}\,\Theta\,\epsilon_\varphi \,S^{(1)}+2\,\varphi'_{0}\,S^{\left(3\right)}+\frac{\varphi'_{0}}{\mathcal{H}}\,S'^{\left(3\right)}+\frac{\varphi'_{0}}{\mathcal{H}}\,S^{\left(2\right)}\,,
\end{equation}
resulting from eq.~\eqref{eqmoto:u:sorgenti}, going back to the variable $\delta\varphi_{\rm flat}$, and accounting only for the dominant term in the source $S_\sigma$. Note that, given \eqref{eq:E:B}, we can approximate
\begin{eqnarray}
S^{(1)}&\simeq& a^2 I\,I_\sigma \,(E_i*E_i+2\,\gamma\, E_i*B_i)\\
S^{(2)}&\simeq& -\frac{I^2\,a^2}{4\,M_{pl}^2}\,E_i*E_i\,.
\end{eqnarray}


\section{The curvature perturbation $\mathcal{R}$}%
\label{sec:curvature}

In the case of single scalar field inflation, solving eq.~\eqref{eqmoto:phi:approx} would be a sufficient step to evaluate the comoving curvature perturbation, which in this case takes the simple form $\mathcal{R}=(\mathcal{H}/\varphi_0')\,\delta\varphi_{\rm flat}$ \cite{Malik:2008im}. 

However, in the presence of both the auxiliary scalar field $\sigma$ and the gauge field $A_\mu$, it is not granted that the above expression for $\mathcal{R}$ holds. We show that, although not exact, the expression $\mathcal{R}=(\mathcal{H}/\varphi_0')\,\delta\varphi_{\rm flat}$ at the end of inflation provides a good approximation in this scenario. Here, again, we follow the notations of \cite{Malik:2008im}. Let us start with the curvature perturbation in uniform density gauge, which can be related to the energy density perturbation in flat gauge as 
\begin{equation}
	\zeta=-\frac{\mathcal{H}}{\rho'_{0}}\,\delta\rho_{\rm flat}
\end{equation}
The derivative of the background energy density $\rho_{0}$ is given by:
\begin{equation}\label{background:energy}
		\rho'_{0}=-3\,\mathcal{H}\,\left(\rho_{0}+P_{0}\right)\simeq -\frac{3\,\mathcal{H}}{a^{2}}\,\left({\varphi'_{0}}^2 +\sigma_0'{}^2\right)\,,
\end{equation}
where $P_{0}$ denotes the pressure. Note that the energy density due to the gauge field is a first order quantity, and therefore it does not contribute to the background energy density. In order to express $\delta\rho_{\rm flat}$ in terms of the fields, we use Einstein equations perturbed at the first order, in particular we use the $0-0$ and $0-i$ equations. Defining metric perturbations in the flat gauge by $\delta g_{00}=-2\,a^{2}\,\Phi_{\rm flat}$ and $\delta g_{0i}=a^{2}\,\partial_i B$, Einstein equations take the form \cite{Malik:2008im}
\begin{align}
&3\,\mathcal{H}^{2}\,\Phi_{\rm flat}-\mathcal{H}\,k^2\,B_{\rm flat}=-4\pi\, G a^{2}\,\delta\rho_{\rm flat}\,,\\
&\mathcal{H}\,\Phi_{\rm flat}-\frac{\varphi'_{0}}{2\,M_{pl}^{2}}\,\delta\varphi_{\rm flat}-\frac{\sigma'_{0}}{2\,M_{pl}^{2}}\,\delta\sigma_{\rm flat}=S^{\left(3\right)}\,.
\end{align}
At large scales the term $k^2\,B_{\rm flat}$ can be neglected, then from the second equation we get the expression of $\Phi_{\rm flat}$, we substitute it in the first equation, and then we get the sought expression for $\delta\rho_{\rm flat}$:
\begin{equation}\label{delta:rho:flat}
	\delta\rho_{\rm flat}=-\frac{3\,\mathcal{H}}{a^{2}}\,\varphi'_{0}\,\delta\varphi_{\rm flat}-\frac{3\,\mathcal{H}}{a^{2}}\,\sigma_0'\,\delta\sigma_{\rm flat}-\frac{6\,\mathcal{H}\,M_{pl}^{2}}{a^{2}}\,S^{\left(3\right)}\,.
\end{equation}
At large scales $\zeta\simeq -\mathcal{R}$ \cite{Malik:2008im}, where $\mathcal{R}$ is the comoving curvature perturbation we are aiming at calculating. Combining \eqref{background:energy} and \eqref{delta:rho:flat}, we obtain the expression of $\mathcal{R}$ in terms of the fields perturbations
\footnote{In principle one should include in eq.~(\ref{delta:rho:flat}) above also terms that are quadratic in the scalar fields. Using the language of the in-in formalism, the inclusion of such terms will be equivalent to considering, in the bispectrum, diagrams with one quartic vertex ($\sim \delta\varphi^2\,A^2$), one cubic vertex  ($\sim \delta\varphi\,A^2$), and two internal photon lines (these diagrams are described in details, albeit is a somehow different model, in~\cite{Carney:2012pk}). On the other hand, the terms proportional to the linear perturbation in $\delta\varphi$ will contribute to the bispectrum a term with three cubic vertices and three internal photon lines. Since each internal photon line carries a contribution~$\propto e^{2\pi\xi}$ that is exponentially large, the terms proportional to the quadratic perturbation in $\delta\varphi$ in eq.~(\ref{delta:rho:flat}) will give a contribution to $f_{NL}$ that is a factor $e^{-2\pi\xi}$ smaller than that from terms that are linear in $\delta\varphi$ and will therefore be neglected in what follows.

Without resorting to the in-in formalism, this result can also be explained as follows. Let us account for these terms schematically as ${\cal R}\sim \delta\varphi_v+\delta\varphi_s+\delta\varphi_v\,\delta\varphi_s$, where $\delta\varphi_v$ denotes the fluctuations in $\varphi$ that are generated by the de Sitter expansion (and that would exist also if the gauge field vanished), and where  $\delta\varphi_s$ denotes the components of $\delta\varphi$ that is sourced by the gauge field at leading order in equation~(\ref{eqmoto:phi:approx}). We are neglecting terms $\sim\delta\varphi_s^2$ as they are higher order in the source. The term proportional to $\delta\varphi_v\,\delta\varphi_s$ contributes both to the two point function and to the three point function. In the two point function  it gives a cross term $\langle {\cal R}\,{\cal R}\rangle\supset\langle(\delta\varphi_v)\,\left(\delta\varphi_v\,\delta\varphi_s\right)\rangle\sim \langle\delta\varphi_v^2\rangle\,\langle\delta\varphi_s\rangle$.  In the three-point function it contributes a term $\langle{\cal R}\,{\cal R}\,{\cal R}\rangle\supset \langle (\delta\varphi_v)\,(\delta\varphi_s)\,(\delta\varphi_v\,\delta\varphi_s)\rangle\sim  \langle \delta\varphi_v^2\rangle\,\langle\delta\varphi_s^2\rangle$. Since $\delta\varphi_s\propto E^2\propto e^{2\pi\xi}$, however, those terms are subdominant with respect to those originating from the linear contribution $\delta\varphi_s$. In fact, the terms in ${\cal R}$ that are linear in $\delta\varphi_s$  give contributions of the order of $e^{4\pi\xi}$ to the two point function and of the order of  $e^{6\pi\xi}$ to the three point function, whereas the term $\sim\delta\varphi_v\,\delta\varphi_s$ gives a contribution $\sim\langle\delta\varphi_s\rangle\sim e^{2\pi\xi}$ to the spectrum and $\sim\langle\delta\varphi_s^2\rangle\sim e^{4\pi\xi}$ to the bispectrum. }
\begin{align}\label{Rfull}
	\mathcal{R}=\frac{\mathcal{H}}{\varphi'_{0}{}^2+\sigma_0'{}^2}\,\left(\varphi_0'\,\delta\varphi_{\rm flat}+\sigma_0'\,\delta\sigma_{\rm flat}+2\,M_{pl}^2\,S^{\left(3\right)}\right)\,.
\end{align}
As expected, the comoving curvature perturbation $\mathcal{R}$ gets contributions directly from each field involved in the Lagrangian eq.~\eqref{lagrangian}.

\paragraph{The term proportional to $S^{\left(3\right)}$.}  Let us first show that the term proportional to $S^{\left(3\right)}$ in eq.~(\ref{Rfull}) can be neglected.  We do so by determining the scale dependence of the following contribution to the curvature power spectrum from Eq.~\eqref{Rfull}
\begin{equation}\label{s3:spectrum}
	\biggl\langle \frac{S^{\left(3\right)}}{\mathcal{H}\,\epsilon_\varphi\,(1+\Theta^2)}\left(\textbf{k}_{1},\tau\right)\frac{S^{\left(3\right)}}{\mathcal{H}\,\epsilon_\varphi\,(1+\Theta^2)}\left(\textbf{k}_{2},\tau\right)\biggr\rangle\,,
\end{equation}
\noindent
with the aim of comparing it with the contribution from the other terms. We define the power spectrum of a generic function $g$ as:
\begin{equation}\label{eq:PSgenerico}
	\langle g\left(\textbf{k}_{1}\right)g\left(\textbf{k}_{2}\right) \rangle = 2\pi^{2}\,\frac{\delta\left(\textbf{k}_{1}+\textbf{k}_{2}\right)}{k_{1}^{3}}\,\mathcal{P}_{g}\left(k_{1}\right)\,,
\end{equation}
and we quantize as above
\begin{equation}\label{quantization:A}
	A_{i}\left(\textbf{x},\tau\right)=\int \frac{{\rm d}^{3}\textbf{p}}{\left(2\pi\right)^{3/2}}e^{i\textbf{p}\cdot\textbf{x}}\varepsilon^{+}_{i}\left(\textbf{p}\right)\left[\hat{a}\left(\textbf{p}\right)+\hat{a}^{\dag}\left(-\textbf{p}\right)\right]\,A_{+}\left(p,\tau\right),
\end{equation}
where we have kept only the positive helicity mode and we have used the fact that the mode functions $A_{+}\left(p,\tau\right)$ are real in the regime $|p\,\tau|\lesssim \xi$ of interest. From eq.~\eqref{eq:S3}, the expression of $S^{\left(3\right)}$ as a function of the gauge field reads
\begin{align}\label{s3:mode}
S^{\left(3\right)}\left(\textbf{k}\right)=\frac{I^{2}}{2\,a^{2}\,M_{pl}^{2}}\int&\frac{{\rm d}^{3}\textbf{p}}{\left(2\pi\right)^{3/2}}\frac{1}{k^{2}}\left[\varepsilon^{+}_{j}\left(\textbf{p}\right)\varepsilon^{+}_{j}\left(\textbf{k}-\textbf{p}\right)\,p_{i}\,k_{i}-p_{j}\,\varepsilon^{+}_{j}\left(\textbf{k}-\textbf{p}\right)\varepsilon^{+}_{i}\left(\textbf{p}\right)\,k_{i}\right]\cdot \nonumber\\
&\left[\hat{a}\left(\textbf{p}\right)+\hat{a}^{\dag}\left(-\textbf{p}\right)\right]\left[\hat{a}\left(\textbf{k}-\textbf{p}\right)+\hat{a}^{\dag}\left(\textbf{p}-\textbf{k}\right)\right]
A_{+}\left(p\right)A_{+}'\left(\left|\textbf{k}-\textbf{p}\right|\right)\,.
\end{align}
By performing some changes of variable and by using the relation (c.f. Eq.~(C3) of \cite{Barnaby:2012xt}) 
\begin{equation}\label{formula}
	k_{i}\,k_{j}\,\varepsilon^{+}_{i}\left(\textbf{p}\right)\varepsilon^{+}_{j}\left(\textbf{k}-\textbf{p}\right)=\left[p\left|\textbf{k}-\textbf{p}\right|+\left(\textbf{k}-\textbf{p}\right)\cdot\textbf{p}\right]\varepsilon^{+}_{i}\left(\textbf{p}\right)\varepsilon^{+}_{i}\left(\textbf{k}-\textbf{p}\right)\,,
\end{equation}
we can rewrite the expression \eqref{s3:mode} as
\begin{align}\label{s3:new}
	S^{\left(3\right)}\left({\mathbf k}\right)=\frac{I^{2}}{2\,a^{2}\,M_{pl}^{2}}\int&\frac{d^{3}{\mathbf p}}{\left(2\pi\right)^{3/2}}\frac{1}{k^{2}}\left[\varepsilon^{+}_{i}\left(\textbf{k}-\textbf{p}\right)\varepsilon^{+}_{i}\left(\textbf{p}\right)\right]
\left[p^{2}-p\left|\textbf{k}-\textbf{p}\right|\right] \cdot \nonumber \\
&\left[\hat{a}\left(\textbf{p}\right)+\hat{a}^{\dag}\left(-\textbf{p}\right)\right]\left[\hat{a}\left(\textbf{k}-\textbf{p}\right)+\hat{a}^{\dag}\left(\textbf{p}-\textbf{k}\right)\right]A_{+}\left(p\right)A_{+}'\left(\left|\textbf{k}-\textbf{p}\right|\right)\,.
\end{align}
From the previous expression, and exploiting the commutation rules of $\hat{a}$ and $\hat{a}^{\dag}$, the power spectrum of the $S^{(3)}/(\mathcal{H}\epsilon_\varphi\,(1+\Theta^2))$ term can be written as (see Eq.~\eqref{eq:PSgenerico})
\begin{align}
{\cal P}_{S^{(3)}}(k)&=\frac{I^{4}\,\mathcal{H}^2}{2\pi^2 \,a^{4}\left(\tau\right)(\varphi_0')^4\,(1+\Theta^2)^2 }\,\int\frac{d^{3}\textbf{p}}{\left(2\pi\right)^{3}}\, \frac{p\left|\textbf{p}-\textbf{k}\right|}{k}\left(p-\left|\textbf{k}-\textbf{p}\right|\right)^{2}\cdot\nonumber\\
&\varepsilon^{+}_{j}\left(\textbf{p}\right)\varepsilon^{+}_{j}\left(\textbf{k}-\textbf{p}\right)\,\varepsilon^{+}_{i}\left(\textbf{p}-\textbf{k}\right)\varepsilon^{+}_{i}\left(-\textbf{p}\right)\,A_{+}\left(p\right)A_{+}'\left(\left|\textbf{k}-\textbf{p}\right|\right)A_{+}\left(\left|\textbf{p}-\textbf{k}\right|\right)A_{+}'\left(-p\right)\,.
\end{align}
We are interested in the behaviour at large scales, so we perform the limit $k\rightarrow 0$, where
\begin{equation}
	\frac{p\left|\textbf{p}-\textbf{k}\right|}{k}\left(p-\left|\textbf{k}-\textbf{p}\right|\right)^{2} \sim k\,p^{2}\,.
\end{equation}
Then, in this limit the spectrum of $S^{(3)}/(\mathcal{H}\epsilon_\varphi\,(1+\Theta^2))$ behaves as:
\begin{equation}
	{\cal P}_{S^{(3)}}(k)\sim \frac{I^{4}\,\mathcal{H}^2}{a^{4}\left(\tau\right)(\varphi_0')^4\,(1+\Theta^2)^2 }\,k\,\int {\rm d}p\,p^{4}\,A_{+}^{2}\left(p\right){A_{+}'}^2\left(p\right)\,.
\end{equation}
The contribution to the curvature power spectrum due to the spectrum of $S^{(3)}/(\mathcal{H}\epsilon\,(1+\Theta^2)) $ (see Eq.~\eqref{Rfull}) is therefore blue, with a spectral index equal to one. We will see in the next section that the contribution to the curvature power spectrum due to the term proportional to $\delta\varphi_{\rm flat}$ is instead scale invariant. The former contribution is therefore largely suppressed at large scales with respect to latter. Note that, from the above result, we can predict that also the contributions due to the correlation between the term $S^{\left(3\right)}/(\mathcal{H}\epsilon\,(1+\Theta^2) )$ and that proportional to $\delta\varphi_{\rm flat}$ are suppressed at large scales and can be neglected.

\paragraph{The term proportional to $\delta\sigma_{\rm {flat}}$.}

We will next assume that the field $\sigma$ gets stabilized at the bottom of its potential  before the end of inflation (this is the situation considered e.g. in~\cite{Ferreira:2014zia,Namba:2015gja}). In this case one should account for the fact that the intensity of the magnetic field will start redshifting away as $a^{-2}$ since the moment in which $\sigma$ has stopped rolling. Therefore one has to assume that $\sigma$ stops rolling towards the end of inflation in order to avoid excessive dilution of the magnetic field during inflation. 

Since $\sigma'_0=0$ at the end of inflation, the curvature perturbation at that stage will be simply given by the single field expression
\begin{equation}\label{R:approx}
	\mathcal{R}\simeq \frac{\mathcal{H}}{\varphi'_{0}}\delta\varphi_{\rm flat}\,
\end{equation}

One should note that in this scenario $\Theta\equiv |\dot\sigma/\dot\varphi|$ will go from large to small and (eventually) vanishing. As we will see below, $\Theta\gg 1$ is needed to require that the beckreaction of the photons on the dynamics of $\sigma$ is negligible during the time in which the relevant scales leave the horizon. When $\sigma$ stabilizes to its minimum and $\Theta$ decreases, however, such backreaction will not be negligible any more, and the presence of a bath of photons will strongly affect the dynamics of $\sigma$ as studied e.g. in~\cite{Anber:2009ua}. Since this will occur when the scales relevant for cosmological magnetic fields are well outside the horizon, we do not expect this complicated dynamics to affect our main results.


\section{Sourced curvature power spectrum and bispectrum}%

In this section we compute the contributions to the scalar spectrum and bispectrum from the excited gauge modes.
 
\subsection{Curvature power spectrum}
\label{sec:spectrum}

Since a source term is present in the equation of motion~\eqref{eqmoto:phi:approx} of $\delta\varphi$, one has
\begin{align}
\delta\varphi=\delta\varphi_{\rm vacuum}+\delta\varphi_{\rm sourced}
\end{align}
where $\delta\varphi_{\rm vacuum}$ is given by the solution to the homogeneous part of eq.~\eqref{eqmoto:phi:approx} with Bunch-Davies boundary condition, whereas $\delta\varphi_{\rm sourced}$ is the particular solution of eq.~\eqref{eqmoto:phi:approx} obtained with a retarded propagator. Since these two components are uncorrelated, the power spectrum $\left\langle \delta\varphi\,\delta\varphi\right\rangle$ receives two contributions: 
\begin{equation}
	\left\langle \delta\varphi\,\delta\varphi\right\rangle=\left\langle \delta\varphi_{\rm vacuum}\,\delta\varphi_{\rm vacuum}\right\rangle+\left\langle \delta\varphi_{\rm sourced}\,\delta\varphi_{\rm sourced}\right\rangle\,,
\end{equation}
where $\left\langle \delta\varphi_{\rm vacuum}\,\delta\varphi_{\rm vacuum}\right\rangle\propto \left(H/2\pi\right)^2$ is the usual contribution obtained from vacuum fluctuations also in the case of single-field slow-roll inflation. Here we calculate the contribution due to the presence of the source term.

As we show in Appendix~\ref{app:simpl_maxwell}, the gravitationally induced part of the source term of eq.~\eqref{eqmoto:phi:approx} can be dramatically simplified by using the equations of motion of the gauge field:
\begin{align}
2\,\varphi'_{0}\,S^{\left(3\right)}+\frac{\varphi'_{0}}{\mathcal{H}}\,S'^{\left(3\right)}+\frac{\varphi'_{0}}{\mathcal{H}}\,S^{\left(2\right)}\simeq \frac{I^2\,\varphi'_{0}\,a^2}{2\,\mathcal{H}\,M_{pl}^2}\,\hat{k}_i\,\hat{k}_j\,E_i*E_j
\,,
\end{align}
where the $\simeq$ sign is due to the fact that we have neglected terms proportional to $B^2$ that, as we have seen above in Eq.~\eqref{eq:E:B}, are subdominant with respect to those proportional to $E^2$ in the regime $\xi\gg 1$ that we are considering.

In terms of $u_{\varphi}=a\,\delta\varphi_{\rm flat}$, the equation of motion~\eqref{eqmoto:phi:approx} takes then the form\footnote{Reference~\cite{Barnaby:2012xt} considered only the second term on the right hand side of eq.~\eqref{eq:moto:u}, whereas~\cite{Ferreira:2014zia} effectively considered only the first one. The first term indeed dominates, as we show in the following. Note that those analyses were performed in the case in which only the $\tilde{F}\,F$ term was coupled to $\sigma$, that corresponds to the case $n\to 0$, $\xi=$constant. Here we perform the full analysis of eq.~\eqref{eq:moto:u} in our model that allows for all values of $-2<n<0$.}
\begin{equation}\label{eq:moto:u}
	u_{\varphi}''+\left(k^{2}-\frac{2}{\tau^{2}}\right)u_{\varphi}= \frac{I^2\,\varphi'_{0}\,a^3}{2\,\mathcal{H}\,M_{pl}^2}\left[2\,N_{\sigma}\,n\,(E_i*E_i+2\,\gamma\, E_i*B_i)+\hat{k}_i\,\hat{k}_j\,E_i*E_j\right]\,,
\end{equation}
where we have used $I_\sigma\simeq n\,\mathcal{H}\,I/\sigma'_0$. Now we quantize the gauge field according to eq.~\eqref{quantization:A} so that the source term above,
\begin{equation}
	J_\bk\left(\tau\right)\equiv \frac{I^2\,\varphi'_{0}\,a^3}{2\,\mathcal{H}\,M_{pl}^2}\left[2\,N_{\sigma}\,n\,(E_i*E_i+2\,\gamma\, E_i*B_i)+\hat{k}_i\,\hat{k}_j\,E_i*E_j\right]
\end{equation}
takes the form
\begin{align}\label{jk}
J_{\textbf{k}}\left(\tau\right)=&\frac{I^2\,\varphi'_{0}}{2\,\mathcal{H}\,M_{pl}^2\,a}\,\int \frac{{\rm d}^{3}\textbf{p}}{\left(2\pi\right)^{3/2}}\varepsilon^+_{i}\left(\textbf{p}\right)\varepsilon^+_{i}\left(\textbf{k}-\textbf{p}\right)\left[\hat{a}_+\left(\textbf{p}\right)+\hat{a}_+^{\dag}\left(-\textbf{p}\right)\right]\left[\hat{a}_+\left(\textbf{k}-\textbf{p}\right)+\hat{a}_+^{\dag}\left(\textbf{p}-\textbf{k}\right)\right] \nonumber\\
&\cdot\left\{\left[2\,n\,N_{\sigma}-\frac{1}{k^{2}}\left[p\left|\textbf{k}-\textbf{p}\right|+\left(\textbf{k}-\textbf{p}\right)\cdot \textbf{p}\right]\right]
A_+'\left(p,\tau\right)A_+'\left(\left|\textbf{k}-\textbf{p}\right|, \tau\right)\right.\nonumber\\
&\quad +4\,N_{\sigma}\,\xi \left|\textbf{k}-\textbf{p}\right|A_+'\left(p,\tau\right)A_+\left(\left|\textbf{k}-\textbf{p}\right|,\tau\right)\Big\}\,.
\end{align}
where we have used the relation~\eqref{formula} and in the last line we have exchanged ${\bf k}-{\bf p}$ with ${\bf p}$, since the integral is symmetric under this exchange. We are now in position to evaluate the sourced scalar spectrum
\begin{eqnarray}
	\lefteqn{\langle \delta\varphi_{\rm sourced}\left({\textbf{k}_{1}},\,\tau\right)\,\delta\varphi_{\rm sourced}\left({\textbf{k}_{2}},\,\tau\right)\rangle=}	\nonumber \\& &\frac{1}{a^{2}\left(\tau\right)}\int_{\tau_{0}}^{\tau}d\tau' G_{\textbf{k}_{1}}\left(\tau,\tau'\right)\int_{\tau_{0}}^{\tau}d\tau'' G_{\textbf{k}_{2}}\left(\tau,\tau''\right)\left\langle J_{\textbf{k}_{1}}\left(\tau'\right)J_{\textbf{k}_{2}}\left(\tau''\right)\right\rangle\,,
\end{eqnarray}
where the Green function $G_{\textbf{k}}\left(\tau,\tau'\right)$, for $\left|k\,\tau\right|<\left|k\,\tau'\right|\ll 1$ is well approximated by (c.f. Eq.~\eqref{eq:Green})
\begin{equation}\label{green:approx}
	G_{k}\left(\tau,\tau'\right)\simeq -\frac{1}{3}\frac{\tau'^{2}}{\tau}\,.
\end{equation}
Applying the commutation rules for the operators $\hat{a}_+$ and $\hat{a}_+^{\dag}$, and exploiting the equality
\begin{equation}\label{proprieta:pol}
	\left|\varepsilon^+_{i}\left(\textbf{k}+\textbf{p}\right)\varepsilon^+_{i}\left(-\textbf{p}\right)\right|^{2}=\frac{1}{4}\left(1+\frac{\left|\textbf{p}\right|^{2}+\textbf{p}\cdot\textbf{k}}{\left|\textbf{p}\right|\cdot\left|\textbf{k}+\textbf{p}\right|}\right)^{2}\,,
\end{equation}
we obtain (recall definition \eqref{eq:PSgenerico})
\begin{align}\label{eq:pscal_sourced} 
	{\cal P}^{\rm {sourced}}_{\delta\varphi}(k)=&\frac{k^3}{2\,\pi^2}\,\frac{\epsilon_\varphi}{32\,H^2\,M_{pl}^2}\int \frac{{\rm d}^{3}\textbf{p}}{\left(2\pi\right)^{3}}\,\left[1-\frac{\textbf{p}\cdot \left(\textbf{k}-\textbf{p}\right)}{p\left|\textbf{k}-\textbf{p}\right|}\right]^{2}\, \Big|\int_{\tau_0}^{\tau} {\rm d}\tau' \left(-H\tau'\right)^{-2n+3}  \nonumber\\
& \cdot\Big\{\left[2\,N_{\sigma}\,n-\frac{1}{k^{2}}\left[p\left|\textbf{k}-\textbf{p}\right|+\left(\textbf{k}-\textbf{p}\right)\cdot \textbf{p}\right]\right]\,A_+'\left(p,\tau'\right)\,A_+'\left(\left|\textbf{k}-\textbf{p}\right|,\tau'\right)\nonumber\\
&\qquad+4\,N_{\sigma}\,\xi\,\left|\textbf{k}-\textbf{p}\right| A_+'\left(p,\tau'\right)\,A_+\left(\left|\textbf{k}-\textbf{p}\right|,\tau'\right)\Big\}\Big|^{2}\,.
\end{align}
We are interested in the power spectrum evaluated at the end of inflation, so that we set $\tau= 0$. The integrand is dominated by modes for which the particle creation is relevant, i.e. $|k\,\tau'|\sim 1/\xi$. Therefore, as long as $\sigma$ is rolling for these values of $\tau'$, we can safely set $\tau_0\rightarrow -\infty$ since this does not include extra relevant contributions to the integral. This puts us in position to use the explicit expressions for the mode functions of the gauge field
\begin{align}
&{A}_+(k,\,\tau)\simeq (-H\,\tau)^n\,\sqrt{-\frac{2\,\tau}{\pi}}\,e^{\pi\xi}\,K_{-2n-1}\left(\sqrt{-8\,\xi\,k\,\tau}\right)\,,\nonumber\\
&{A}_+(k,\,\tau)'\simeq (-H\,\tau)^n\,\sqrt{\frac{4\,\xi\,k}{\pi}}\,e^{\pi\xi}\,K_{-2n}\left(\sqrt{-8\,\xi\,k\,\tau}\right)\,,
\end{align}
and insert them into eq.~\eqref{eq:pscal_sourced}. We then change variables to $y=\sqrt{-8\,\xi\,k\,\tau}$, choose a reference frame where the $z$-axis is parallel to $\textbf{k}$, and define $Q\equiv p/k$, $\mu\equiv\cos\alpha$, where $\alpha$ is the angle between $\textbf{k}$ and $\textbf{p}$. The power spectrum then reads
\begin{align}\label{spettro:tot}
	{\cal P}^{\rm {sourced}}_{\delta\varphi}(k)=&\,\epsilon_\varphi\,\frac{e^{4\pi\xi}}{\xi^{6}}\frac{H^{4}}{9\times 2^{23}\,\pi^{6}\,M_{pl}^2}\int_{0}^{\infty} dQ\,Q^{3} \int_{-1}^{1} d\mu \,P\left(1-\frac{\mu-Q}{P}\right)^{2} \nonumber \\
	&\cdot\Big|\int_{0}^{\infty}dy\,y^{7}\Big\{\left[2\,N_{\sigma}n-Q\,\left(P+\mu-Q\right)\right]\left[K_{-2n}\left(y\,\sqrt{Q}\right)\,K_{-2n}\left(y\,\sqrt{P}\right)\right]\nonumber\\
	&+N_{\sigma}\,\sqrt{P}\,\left[y\,K_{-2n-1}\left(y\sqrt{P}\right)\,K_{-2n}\left(y\,\sqrt{Q}\right)\right]\Big\}\Big|^{2}\,,
\end{align}
where $P\equiv \sqrt{1+Q^{2}-2\,Q\,\mu}$. The first term proportional to $N_{\sigma}$ is the one due to the first term in \eqref{eq:moto:u}, the term proportional to $Q\left(P+\mu-Q\right)$ corresponds to the part of the source proportional to $\hat{k}_{i}\hat{k}_{j}E_{i}\ast E_{j}$, and the last term to the part of the source proportional to $\gamma\, E_{i} \ast B_{i}$.

\begin{figure}
    \centering	
		\includegraphics[width=0.6\textwidth]{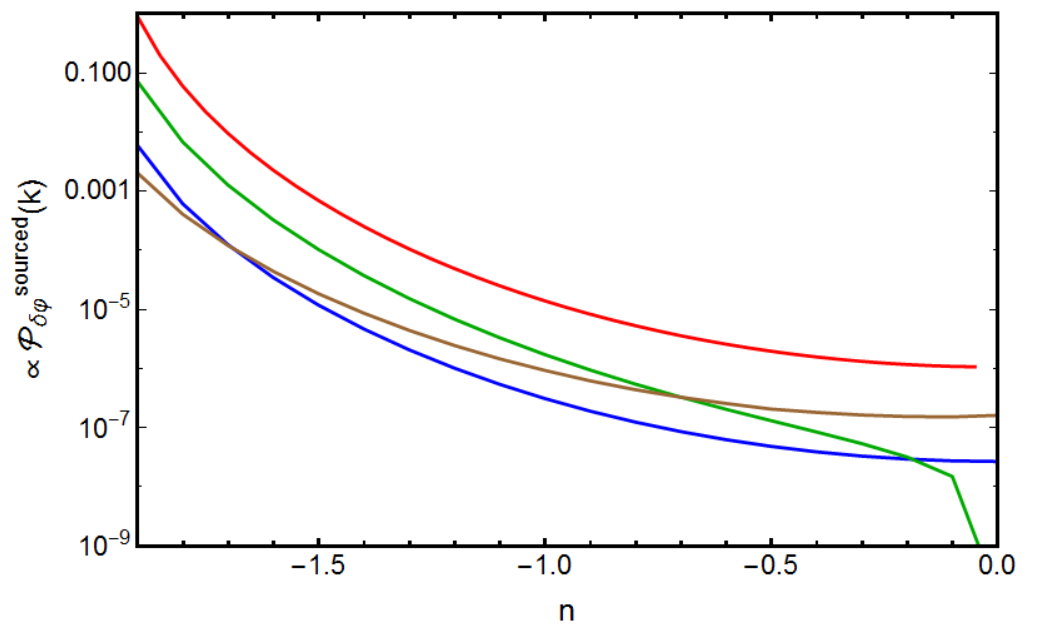}
    \caption{\small Relative contributions to the curvature power spectrum given in Eq.~\eqref{spettro:tot}, where we have fixed $N_\sigma=1$ as the limiting case. Contributions are shown up to the common factor $\epsilon_\varphi\left(n\,\frac{e^{2\pi\xi}}{\xi^{3}}\,\frac{H^{2}}{\pi\,M_{pl}}\right)^2$. From top to bottom, the upper red curve shows the power spectrum due to the sum of the two terms proportional to $N_{\sigma}^{2}$; in green the contribution to the power spectrum due to the cross-correlation between the term proportional to $N_\sigma E_{i}\ast E_{i}$ and the gravitationally induced term proportional to $\hat{k}_i\hat{k}_jE_i\ast E_j$; in brown the contribution due to the cross-correlation between the term proportional to $N_\sigma\gamma E_{i}\ast B_{i}$ and the gravitationally induced term; in blue the power spectrum due to the gravitationally induced term.}
    \label{spettri}
\end{figure}

In Figure~\ref{spettri}, we show the relative contributions to the power spectrum due to each term in \eqref{spettro:tot}, collecting in a unique quantity the two terms proportional to $N_{\sigma}^{2}$. Contributions are shown up to a common factor.

It appears from Fig.~\ref{spettri} that the contributions to the power spectrum which are not proportional to $N_{\sigma}^{2}$ are at least one order of magnitude smaller than the other ones already for $N_{\sigma}=1$. Therefore, we can safely approximate the source of the inflaton perturbations by the first two terms in Eq.~\eqref{eq:moto:u}. We use this fact in the next section to simplify the calculation of the bispectrum. The results we obtain for the curvature power spectrum and bispectrum sourced by the gauge field are accurate and can be compared to current observations. 


\subsection{Curvature bispectrum}
\label{sec:bispectrum}

For the calculation of the curvature bispectrum we proceed analogously to the power spectrum. As before, the amplitude of the bispectrum of the inflaton perturbations is given by the standard vacuum contribution plus the one due to the presence of the gauge field which represent active sources. The former is well known to be very small, and here we calculate the latter.

As before, we refer to the equation of motion for $u_{\varphi}$, eq.~\eqref{eq:moto:u}. Using the expression for the source term~\eqref{jk} where we suppress the subdominant, gravitationally induced term not proportional to $N_\sigma$, and the Green function~\eqref{green:approx}, the three point function of $\delta\varphi_{\rm sourced}$ takes the form
\begin{align}\label{bispettro1}
	\left\langle \delta\varphi_{\textbf{k}_{1}}\delta\varphi_{\textbf{k}_{2}}\delta\varphi_{\textbf{k}_{3}}\right\rangle=&\frac{n^3}{a^{3}}\frac{N_{\sigma}^{3}\epsilon_\varphi^{3}}{\dot{\varphi}^{3}}\frac{e^{6\pi\xi}}{\xi^{9}}\frac{1}{9\times 2^{24}\pi^{3}} \delta\left(\textbf{k}_{1}+\textbf{k}_{2}+\textbf{k}_{3}\right)\cdot\int\frac{{\rm d}^{3}\textbf{p}}{\left(2\pi\right)^{9/2}} \nonumber\\
	&\cdot\left[\varepsilon^{+}_{i}\left(\textbf{p}\right)\varepsilon^{+}_{i}\left(\textbf{p}-\textbf{k}_{1}\right)\right]
	\left[\varepsilon^{+}_{j}\left(\textbf{p}-\textbf{k}_{1}\right)\varepsilon^{+}_{j}\left(-\textbf{k}_{3}-\textbf{p}\right)\right]
	\left[\varepsilon^{+}_{l}\left(\textbf{p}+\textbf{k}_{3}\right)\varepsilon^{+}_{l}\left(-\textbf{p}\right)\right] \nonumber\\
	&\int_{0}^{\infty} {\rm d}x' h\left(\left|\textbf{p}\right|,\left|\textbf{k}_{1}-\textbf{p}\right|;x'\right)\int_{0}^{\infty}  {\rm d}x'' h\left(\left|\textbf{p}-\textbf{k}_{1}\right|,\left|\textbf{k}_{3}+\textbf{p}\right|;x''\right) \nonumber \\
	&\int_{0}^{\infty}  {\rm d}x''' h\left(\left|\textbf{p}+\textbf{k}_{3}\right|,\left|\textbf{p}\right|;x'''\right)\,,
\end{align}
where we have defined
\begin{align}
	h\left(\left|\textbf{p}\right|,\left|\textbf{k}_{1}-\textbf{p}\right|;x\right)=&\frac{\sqrt{\left|\textbf{p}\right|\cdot\left|\textbf{k}_{1}-\textbf{p}\right|}}{H}\, x^{7}\,\Bigg[K_{-2n}\left(x\sqrt{\frac{\left|\textbf{k}_{1}-\textbf{p}\right|}{H}}\right)K_{-2n}\left(x\sqrt{\frac{\left|\textbf{p}\right|}{H}}\right)\nonumber\\
	&-\frac{x}{2n}\frac{\sqrt{\left|\textbf{k}_{1}-\textbf{p}\right|}}{\sqrt{H}}K_{-2n}\left(x\sqrt{\frac{\left|\textbf{p}\right|}{H}}\right)K_{-2n-1}\left(x\sqrt{\frac{\left|\textbf{k}_{1}-\textbf{p}\right|}{H}}\right)\Bigg]\,.
\end{align}
Exploiting the relation \eqref{proprieta:pol}, the products of the polarization vectors in the second line of eq.~\eqref{bispettro1} can be written as
\begin{equation}
\begin{split}
	\left[\varepsilon\mbox{ product}\right]&\equiv \frac{1}{8}\big[\left(\hat{q}_{1}\cdot\hat{q}_{3}\right)^{2}+
	\left(\hat{p}\cdot\hat{q}_{3}\right)^{2}+
	\left(\hat{p}\cdot\hat{q}_{1}\right)^{2}+
	\left(\hat{p}_{1}\cdot\hat{q}_{3}\right)+
	\left(\hat{p}\cdot\hat{q}_{3}\right)+
	\left(\hat{p}_{1}\cdot\hat{p}\right)+ \\
	&\left(\hat{q}_{1}\cdot\hat{p}\right)\left(\hat{p}\cdot\hat{q}_{3}\right)+\left(\hat{p}\cdot\hat{q}_{1}\right)\left(\hat{q}_{1}\cdot\hat{q}_{3}\right)+
	\left(\hat{p}\cdot\hat{q}_{3}\right)\left(\hat{q}_{1}\cdot\hat{q}_{3}\right)-
	\left(\hat{p}\cdot \hat{q}_{1}\right)\left(\hat{p}\cdot\hat{q}_{3}\right)\left(\hat{q}_{1}\cdot\hat{q}_{3}\right)\big]
	\end{split}
\end{equation}
where $\textbf{q}_{1}=\textbf{p}-\textbf{k}_{1}$ and $\textbf{q}_{3}=\textbf{p}+\textbf{k}_{3}$.

In scenarios where non-gaussianities are generated by subhorizon dynamics, we expect the bispectrum to be maximized in the equilateral configuration $k\equiv \left|\textbf{k}_{1}\right|=\left| \textbf{k}_{2}\right|=\left|\textbf{k}_{3}\right|$. In our model this is the case if $n$ is sufficiently different from $-2$, so that the energy in the gauge field drops rapidly at superhorizon scales. In the limit $n\to -2$ the scale invariant electric field will keep sourcing $\delta\varphi$ outside the horizon, and we expect the bispectrum to converge to the local shape.  Nevertheless, in what follows we will estimate the magnitude of the bispectrum by computing the three-point function in the equilateral configuration, that we expect to provide a reasonable estimate of the nongaussianities in the model for values $-2<n<0$. We obtain
\begin{equation}\label{bispettro}
	\left\langle \delta\varphi_{\textbf{k}_{1}}\delta\varphi_{\textbf{k}_{2}}\delta\varphi_{\textbf{k}_{3}}\right\rangle_{\rm equil}=\frac{\delta\left(\textbf{k}_{1}+\textbf{k}_{2}+\textbf{k}_{3}\right)}{k^{6}}\frac{H^{6}}{M_{pl}^{3}}\frac{e^{6\pi\xi}}{\xi^{9}}\left(N_{\sigma}\,\sqrt{2\epsilon_\varphi}\right)^{3}f\left(n\right)\,,
\end{equation}
where
\begin{align}\label{eq:def_fn}
	&f\left(n\right)=\frac{n^3}{9\times 2^{27}\,\pi^{3}}\frac{1}{\left(2\pi\right)^{9/2}}\int_{0}^{\infty}  dP\,P^{3}\int_{-1}^{1} d\cos\bar{\Theta}\,Q\,R\cdot\left[\varepsilon \mbox{ product}\right]\cdot \nonumber\\
	&\cdot\int_{0}^{\infty} dy_{1}\, y_{1}^{7}\left[K_{-2n}\left(y_{1}\sqrt{Q}\right)K_{-2n}\left(y_{1}\sqrt{P}\right)-\frac{y_{1}}{2n}\sqrt{Q}K_{-2n}\left(y_{1}\sqrt{P}\right)K_{-2n-1}\left(y_{1}\sqrt{Q}\right)\right]\nonumber\\
	&\cdot\int_{0}^{\infty} dy_{2}\, y_{2}^{7}\left[K_{-2n}\left(y_{2}\sqrt{Q}\right)K_{-2n}\left(y_{2}\sqrt{R}\right)-\frac{y_{2}}{2n}\sqrt{R}K_{-2n}\left(y_{2}\sqrt{Q}\right)K_{-2n-1}\left(y_{2}\sqrt{R}\right)\right] \nonumber\\
	&\cdot\int_{0}^{\infty} dy_{3}\, y_{3}^{7}\left[K_{-2n}\left(y_{3}\sqrt{R}\right)K_{-2n}\left(y_{3}\sqrt{P}\right)-\frac{y_{3}}{2n}\sqrt{P}K_{-2n}\left(y_{3}\sqrt{R}\right)K_{-2n-1}\left(y_{3}\sqrt{P}\right)\right]\,,
\end{align}
where we have aligned $\bk_1$ along the third axis, and we have defined $P\equiv p/k_{1}$, $\bar{\Theta}$ is the angle between $\textbf{p}$ and $\textbf{k}_{1}$, $\bar{\Phi}$ the angle between $\textbf{p}$ and the $x$-direction, and 
\begin{align}
	Q\equiv\sqrt{1-P^{2}-2P\cos\bar{\Theta}}\,,\qquad 
	R\equiv\sqrt{1+P^{2}+2P\left(-\frac{1}{2}\cos\bar{\Theta}+\frac{\sqrt{3}}{4}\,\sin\bar{\Theta}\right)}\,.
\end{align}

As mentioned above, in the following we are going to consider only the result of the equilateral configuration, and compare it with current upper limits on curvature non-gaussianities in terms of $f_{\rm NL}^{\rm equil}$.


\section{Constraints on the magnetic field production from curvature perturbations}

In the previous sections we obtained the curvature power spectrum and bispectrum as a function of the parameters of the model. We now compare them to the results provided by observations, in order to constrain the parameter space. We also require that the lower bound on magnetic fields in the intergalactic medium is satisfied, see e.g.~\cite{Neronov:1900zz,Taylor:2011bn,Vovk:2011aa}. We find that the combination of these requirements can be satisfied: our model can explain the presence of intergalactic magnetic fields and satisfy, at the same time, the constraints imposed by the curvature bispectrum.

Observations provide strict constraints on the curvature non-gaussianities. In particular, for the equilateral configuration, \cite{Ade:2015ava} finds the following bound: $f_{\rm NL}^{\rm equil}=-16\pm 70$ from temperature data, $f_{\rm NL}^{\rm equil}=-3.7\pm 43$ for $T+E$ data, at $68\%$ of CL \cite{Ade:2015ava}.
As usual, we parametrize the bispectrum by $f_{\rm NL}^{\rm equil}$, defined as
\begin{equation}
	\left\langle \mathcal{R}\left(\textbf{k}_{1}\right)\mathcal{R}\left(\textbf{k}_{2}\right)\mathcal{R}\left(\textbf{k}_{3}\right)\right\rangle=\frac{3}{10}\left(2\pi\right)^{5/2}f_{\rm NL}^{\rm equil}\mathcal{P}_{\mathcal{R}}^{2}\delta\left(\textbf{k}_{1}+\textbf{k}_{2}+\textbf{k}_{3}\right)\frac{\sum k_{i}^{3}}{\Pi k_{i}^{3}}\,,
\end{equation}
where $\mathcal{P}_{\mathcal{R}}$ is the total curvature power spectrum.
From the expression of the bispectrum \eqref{bispettro}, using the relation \eqref{R:approx}, we have
\begin{equation}\label{fnl}
	f_{\rm NL}^{\rm equil}=\frac{10}{9}\frac{1}{\left(2\pi\right)^{5/2}}\frac{1}{\mathcal{P}_{\mathcal{R}}^{2}}\frac{H^{6}}{M_{pl}^{6}}\frac{e^{6\pi\xi}}{\xi^{9}}N_{\sigma}^{3}\,f\left(n\right)\,,
\end{equation}
where we used $\dot{\varphi}_0=\sqrt{2\epsilon_\varphi}\,M_{pl}\,H$, and where $f(n)$, which is defined in eq.~(\ref{eq:def_fn}) above, is well approximated by $f(n)\simeq 1.7\times 10^{-5}\,(2+n)^{4.8}$.

We impose $f_{\rm NL}^{\rm equil}$ in \eqref{fnl} to be equal or smaller than the current upper bound. At the same time, we impose sufficient magnetogenesis to satisfy the lower bound on  magnetic fields in the intergalactic medium.

\subsection{Constraints on inflationary parameters}
\label{sec:constraints}

In~\cite{Caprini:2014mja} it was found that the magnetic field produced in this model is given, at the end of inflation, by
\begin{equation}\label{B:reh}
	B_{\rm reh}=H^{2}\frac{e^{\pi\xi}}{\xi^{5/2}}b_{n}\, e^{-2\,\Delta N}\,, \qquad \mbox{where} \qquad b_{n}\equiv \sqrt{\frac{\Gamma\left(4-2n\right)\Gamma\left(6+2n\right)}{80640\, \pi^{3}}}\,,
\end{equation}
and where we have introduced a factor $e^{-2\,\Delta N}$ to account for the dilution of the magnetic field during inflation after the field $\sigma$ has stopped rolling. Since large values of $\Delta N$ will lead to weak magnetic fields, we will assume in what follows that $\Delta N\lesssim 1$. The correlation length evaluates to
\begin{equation}\label{L:reh}
	L_{\rm reh}=\frac{\xi}{H}\,l_{n}\,,  \qquad \mbox{where} \qquad l_{n}=\frac{18\pi}{\left(3-2n\right)\,\left(5+2n\right)}\,.
\end{equation}
The magnetic field at the present time (a subscript ${}_0$ indicates quantities evaluated today) is obtained from the relations \cite{Caprini:2014mja} 
\begin{equation}\label{eqs:inverse_cascade}
	B_{0}=10^{-8}\mbox{G}\,\frac{L_{0}}{{\rm Mpc}}\,, \qquad B_{0}^{2}\,L_{0}=B_{\rm reh}^{2}\,L_{\mathrm {reh}}\left(\frac{a_{\mathrm {reh}}}{a_{0}}\right)^{3}\,,
\end{equation}
that are a consequence of the self-similar evolution associated to the inverse cascade of the helical field~\cite{Banerjee:2004df,Durrer:2013pga,Caprini:2014mja}. The above results~(\ref{eqs:inverse_cascade}) are derived for magnetic fields with a finite correlation length. The effects of the inverse cascade are different (and weaker) for scale invariant fields~\cite{Brandenburg:2016odr,Kahniashvili:2016bkp,Brandenburg:2018ptt}, which correspond to the case $n= -2$. In our analysis we will assume that $n$ is sufficiently far from $-2$, so that eq.~(\ref{eqs:inverse_cascade}) can be assumed to be approximately valid. Therefore we have
\begin{align}\label{B:today}
	&\frac{L_{0}}{{\rm Mpc}}=\left(\frac{a_{\rm reh}}{a_{0}}\right)\,\left(\frac{B_{\rm reh}}{10^{-8}\mbox{G}}\right)^{2/3}\,\left(\frac{L_{\rm reh}}{{\rm Mpc}}\right)^{1/3}\,,\nonumber \\
	& B_{0}=10^{-8}\mbox{G}\,\left(\frac{a_{\rm reh}}{a_{0}}\right)\,\left(\frac{B_{\rm reh}}{10^{-8}\mbox{G}}\right)^{2/3}\left(\frac{L_{\rm reh}}{{\rm Mpc}}\right)^{1/3}\,.
\end{align}
The lower bound on intergalactic magnetic fields from~\cite{Neronov:1900zz,Taylor:2011bn,Vovk:2011aa}, including the correction arising from the dependence on the magnetic field spectral index found in \cite{Caprini:2015gga}, reads
\begin{equation}\label{bnv:constraint}
	B_{0}\,\sqrt{\frac{L_{0}}{{D_e}}}\,\Pi\left(\frac{D_e}{L_0},n_B\right)\geq B^{NV}\,,
\end{equation}
where $B^{NV}\simeq 10^{-18}\div 10^{-16}$~G. Since in this context $D_e\simeq 80$ kpc \cite{Neronov:2009gh}, in the above equation we have selected the appropriate case $L_0\ll D_e$ \cite{Caprini:2014mja}. The function $\Pi\left(\frac{D_e}{L_0},n_B\right)$ takes a particularly simple form if $n_B>1/2$ (corresponding to $n_G>-2$, where $n_G=2n_B-3$ is the spectral index as defined in \cite{Caprini:2015gga}), which is always the case for the range of values $-2<n<0$, as can be verified from Eq.~\eqref{eq:nB}. Given the conditions $L_0\ll D_e$ and $-2<n<0$, it is therefore enough to require that our model leads to a present time magnetic field satisfying the (in-)equality
\begin{equation}\label{bnv:constraintnB}
	B_{0}\,\sqrt{\frac{L_{0}}{{D_e}}}\,\sqrt\frac{n_B}{10n_B-5}\geq B^{NV}\,,
\end{equation}
with $B^{NV}=10^{-17}$~G: this guarantees that the model provides a strong enough intergalactic magnetic field.

By using the first of eqs.~(\ref{eqs:inverse_cascade}) along with the saturated inequality eq.~(\ref{bnv:constraintnB}), with $B^{NV}=10^{-17}$~G one derives the coherence length $L_0\simeq 1$~pc and the amplitude $B_0\simeq 10^{-14}$~G. This implies, in particular, that the intensity of the magnetic field is much smaller than that, of the order of $10^{-9}$~G, that is necessary for the magnetic field to directly generate observable effects in the CMB~\cite{Ballardini:2014jta}.

To proceed further, we use
\begin{equation}
	\frac{a_{0}}{a_{\rm reh}}=\frac{g_{\ast}^{reh\,1/3}T_{\rm reh}}{g_{\ast}^{0\,1/3}T_{0}}\,,
\end{equation}
where $g_{\ast}^{0}=3.36$ and $g_{\ast}^{\rm reh}=106.75$ are the effective number of relativistic degrees of freedom at the respective temperatures, $T_{0}=2\times 10^{-13}$~GeV, and the reheating temperature $T_{\rm reh}$ is obtained assuming instantaneous reheating:
\begin{equation}
	\frac{g_{\ast}^{\rm reh}\,\pi^{2}}{30}\,T_{\rm reh}^{4}=3\,M_{pl}^{2}\,H^{2}\,.
\end{equation}
As a consequence we have $\frac{a_{0}}{a_{\rm reh}}\simeq 2\times 10^{31}\sqrt{\frac{H}{M_{pl}}}$, so that, using eq.~\eqref{bnv:constraintnB}, one gets
\begin{equation}\label{eq:xi_from_bnv}
	\frac{e^{\pi\xi}}{\xi^{2}}=60\,e^{2\,\Delta N}\,\left(\frac{M_{pl}}{H}\right)^{3/4}\,\frac{B^{NV}_{-17}(n_B)}{b_{n}\sqrt{l_{n}}}\,,
\end{equation}
where 
\begin{equation}
	B^{NV}_{-17}(n_B)=\sqrt{\frac{D_e}{\rm Mpc}}\,\sqrt\frac{10n_B-5}{n_B} \frac{B^{NV}}{10^{-17}~{\rm G}}\,.
\end{equation}
Inserting $D_e\simeq 80$ kpc and using relation~\eqref{eq:nB} to express $n_B$ as a function of $n$, Eq.~\eqref{eq:xi_from_bnv} gives us the parameter $\xi$ as a function of the three parameters $\xi=\xi\left(H,n,\Delta N\right)$.

Now we impose the requirement that the bispectrum of the curvature perturbations does not exceed the observational constraints provided by Planck. To determine the boundaries of the allowed parameter space we fix the value of $f_{\rm NL}^{\rm equil}$ to $50$ and $10$. Eq.~\eqref{fnl} with $\mathcal{P}_{\mathcal{R}}=2.21\cdot 10^{-9}$ provides the Hubble parameter as a function of $\xi$, $n$ and $N_{\sigma}$, i.e. $H=H\left(\xi,\,n,\,N_{\sigma}\right)$. Inserting $\xi=\xi\left(H,n,\Delta N\right)$ from Eq.~\eqref{eq:xi_from_bnv} we obtain $H$ as a function of $n$, $N_{\sigma}$ and $\Delta N$. Such a function identifies the region of parameter space that leads to successful magnetogenesis and, at the same time, does not exceed the upper bound on $f_{\rm NL}^{\rm equil}$ provided by Planck.

\begin{figure}
    \centering
		\includegraphics[width=0.7\textwidth]{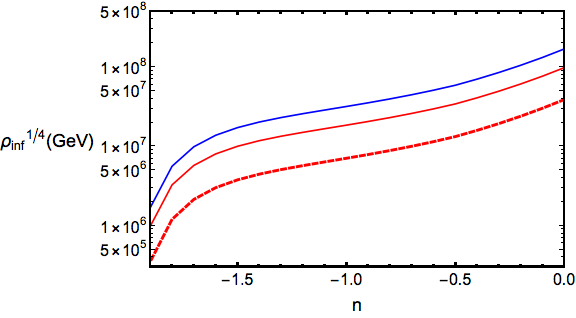}
    \caption{\small Constraints on the inflationary energy density in the model. We have imposed that the magnetic field amplitude today satisfies eq.~\eqref{bnv:constraintnB} with $B^{NV}=10^{-17}$~G.  The top blue curve is obtained by fixing $f_{NL}^{\rm equil}=50$ and $N_\sigma=N_{\rm obs}$, the curve in the middle has $f_{NL}^{\rm equil}=10$, still with $N_\sigma=N_{\rm obs}$. Finally, the curve at the bottom has $f_{NL}^{\rm equil}=50$ with $N_\sigma=200$. In all the curves we have assumed that $\Delta N=0$. The energy scale of inflation quickly decreases as we increase $\Delta N$, as it goes as $e^{-4\,\Delta N}$.}
    \label{plot:constraints}
\end{figure}

To proceed, we note that in principle $N_\sigma$ can be smaller or larger than $N_{\rm obs}$, the number of efoldings of observable inflation. In fact, in order for this mechanism to lead to successful magnetogenesis,  all we need is that $\sigma$ is rolling when the coherence length of the magnetic field (in order to explain the blazar observation) and when galactic scales (in order to explain the galactic field) are leaving the horizon. Both scales are much shorter than COBE scales. However, in this paper we are assuming that $\sigma$ is rolling when COBE scales are leaving the horizon, so that the magnetic field sources nongaussianities that are observable in the CMB. This assumption, together with the fact that $\sigma$ stops rolling only towards to the end of inflation, implies that we cannot assume that $N_\sigma$ is much smaller than $N_{\rm obs}$. In what follows we will therefore assume that 
\begin{equation}
	  N_\sigma\gtrsim N_{\rm obs}={\rm Log}\left[\left(\frac{g_{\ast}^{\rm reh}}{g_{\ast}^{0}}\right)^{1/3}\frac{M_{pl}}{T_{0}}\left(\frac{g_{\ast}^{\rm reh}\pi^{2}}{30}\right)^{1/4} \sqrt{\frac{H}{M_{pl}}}\right]\,.
\end{equation}
Figure~\ref{plot:constraints} shows that the energy scale of inflation spans a few of orders of magnitude $10^{5}\lesssim \rho_{\rm inf}^{1/4} \lesssim 10^{8}$~GeV as\footnote{Note that our findings are consistent with those of~\cite{Fujita:2015iga}, that has shown that the magnetic field produced by the axial coupling is not strong enough to satisfy the blazar constraint if inflation happens at chaotic inflation energy scales $\rho_{\rm inf}^{1/4} \sim 10^{16}$~GeV.}  $n$ varies from $-2$ to $0$. This expression is valid for $\Delta N=0$, and should be rescaled by $\sim e^{-4\,\Delta N}$ for $\Delta N\neq 0$. To see this, let us note that by choosing a fixed value of $f_{NL}$ we are fixing the combination $e^{6\pi\xi}\,H^6/\xi^9$. Using eq.~(\ref{eq:xi_from_bnv}), once $B^{NV}$ is fixed, this is equivalent to  fixing the quantity $H^{3/2}\,e^{12\,\Delta N}\,\xi^3$. Neglecting the (weak) dependence on the factor $\xi^3$, this is equivalent to fixing $\rho_{\rm inf}^{3/4}\,e^{12\,\Delta N}$ so that $\rho_{\rm inf}^{1/4}\propto e^{-4\,\Delta N}$.

The strong dependence of $\rho_{\rm inf}$ in $\Delta N$ implies that $\sigma$ should become inert very close to the end  of inflation. Under this condition, the model provides a possible explanation for the observed presence of magnetic fields in the intergalactic medium, while respecting current observational constraints from scalar modes, and without requiring inflation to happen at energy scales below the TeV or so. 

\begin{figure}
    \centering
		\includegraphics[width=0.6\textwidth]{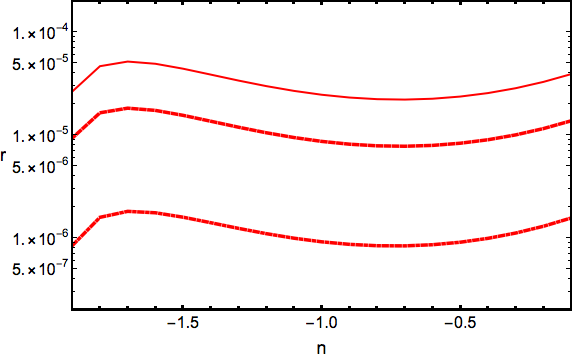}
    \caption{The tensor to scalar ratio $r$. The three curves (top curve to bottom curve) give the value of $r$ as a function of $n$ for the following values of $f_{NL}$ and $N_\sigma$ : $f_{\rm NL}^{\rm equil}=50$ and $N_{\sigma}=N_{\rm obs}$, $f_{\rm NL}^{\rm equil}=10$ and $N_{\sigma}=N_{\rm obs}$, and $f_{\rm NL}^{\rm equil}=50$ and $N_{\sigma}=200$ . }
    \label{fig:tensor}
\end{figure}

\vspace*{0.5cm}
\noindent
Concerning tensor modes, the tensor-to-scalar ratio is given in Eq.~(3.6) of~\cite{Caprini:2014mja}: 
\begin{equation}
	r=p^t \left(n\right)\frac{H^{4}}{M_{pl}^{4}}\frac{e^{4\pi\xi}}{\xi^{6}}\mathcal{P}_\mathcal{R}^{-1}\,,
\end{equation}
where $p^{t}\left(n\right)$ is plotted in Figure 1 of \cite{Caprini:2014mja}. It is straightforward to see that $r$ is proportional to $f_{NL}^{2/3}\,N_\sigma^{-2}$. In  Fig.~\ref{fig:tensor} we show the tensor to scalar ratio for $f_{NL}=50$, $N_\sigma=N_{\rm obs}$, for $f_{NL}=10$, $N_\sigma=N_{\rm obs}$ and for $f_{NL}=50$, $N_\sigma=200$. In all cases the tensor-to-scalar ratio in this model is too small to be detected both by future Earth- and space-based dedicated missions, even if it is not hugely below the projected sensitivity $r\sim 10^{-3}$ of the CMB-S4 experiment~\cite{Abazajian:2016yjj}. As explained in section \ref{sec:tensor}, tensor modes due to vacuum fluctuations of the gravitational field are negligible with respect to those generated by the presence of the gauge field: our model would therefore be further constrained by a future detection of CMB B-polarisation of primordial origin.

Before concluding this Section on the constraints on the model, we discuss under which conditions we can neglect the backreaction of the mode functions of the photon on the dynamics of $\sigma$. Since the energy in $\sigma$ available for particle production is only the kinetic component $\frac{1}{2}\dot\sigma^2$, we require that $\langle\rho_{\rm EM}\rangle=\frac{1}{2}\langle E^2(\bx)+B^2(\bx)\rangle\ll \frac{1}{2}{\dot\sigma_0^2}$.

The energy density in the gauge modes reads
\begin{align}
\langle\rho_{EM}\rangle&=\frac{I^2(\tau)}{2\,a(\tau)^4}\int\frac{d\bk}{(2\pi)^3}\left(A_+'(k,\,\tau)^2+k^2\,A_+(k,\,\tau)^2\right)\simeq H^4\,\frac{e^{2\pi\xi}}{\xi^3}\,n\,\frac{9-n^2(7-4\,n^2)^2}{1120\,\pi^2\,\sin(2\,n\,\pi)}
\end{align}
where in the last equality line we have kept the leading term in the limit $\xi\gg 1$. We this see that $\langle\rho_{EM}\rangle$ takes values of the order of $10^{-3}\,H^4\,\frac{e^{2\pi\xi}}{\xi^3}$ unless $n$ is very close to $-2$, in which case we have $\rho_{EM}(\tau)\simeq \frac{.01}{n+2}\, H^4\,\frac{e^{2\pi\xi}}{\xi^3}$.

Using eq.~(\ref{eq:xi_from_bnv}), we obtain
\begin{align}
\frac{\rho_{EM}}{\dot\sigma^2/2}&\simeq .25\,\frac{e^{4\,\Delta N}}{\epsilon_\sigma}\,\frac{\rho_{\rm inf}^{1/4}}{M_{pl}}\,B^{NV}_{-17}(n_B)^2\,\xi\,n\,\frac{9-n^2(7-4\,n^2)^2}{{b_{n}^2\,{l_{n}}}\,\sin(2\,n\,\pi)}\,.
\end{align}
Using the values of $e^{4\,\Delta N}\,{\rho_{\rm inf}^{1/4}}/{M_{pl}}$ extracted from Figure~\ref{plot:constraints}, along with an evaluation of the $n$-dependent factor on the right hand side of the equation above, and using $\xi\sim 10$, we obtain that ${\rho_{EM}}/(\dot\sigma^2/2)\sim (10^{-7}\div 10^{-8})/\epsilon_\sigma$ so that the backreaction of the gauge modes on the background evolution of $\sigma$ is negligible as long as $\epsilon_\sigma\gtrsim 10^{-7}$, which is well within the slow-roll regime.

\subsection{Possible seeds for the galactic dynamo}

Galaxies appear to be magnetised with magnetic field intensity of the order of $10^{-6}\,{\rm G}$. It is believed that such magnetic fields are the result of the amplification of pre-existing seeds by the galactic dynamo mechanism (see e.g. \cite{Brandenburg:2004jv} and references therein). No preferred explanation exists so far for the origin of the magnetic seed, which could be both astrophysical (see e.g. \cite{Kulsrud:2007an}) or primordial. 
There are some uncertainties on the efficiency of the dynamo, but reliable estimations indicate that the initial magnetic field should be at least of the order of $10^{-23}-10^{-21}{\rm G}$ at a comoving scale of $1\,{\rm Mpc}$, in order to explain the observed $\mu$G amplitude\footnote{Note that Ref.~\cite{Davis:1999bt} has found that in a $\Lambda$CDM universe a much smaller amplitude might be needed, in reason of the longer time-scale of evolution of the galaxies: a seed of $10^{-30}$ G could actually be enough to give rise to $\mu$G field today.} \cite{Brandenburg:2004jv}. 

According to some numerical simulations of the evolution of magnetised structures, once galaxies are magnetised, this could also explain the observed magnetisation of clusters: see e.g. \cite{Donnert:2008sn}. The ejection of the galactic magnetic fields via galactic winds and a subsequent amplification at large scales by intra-cluster turbulence can in principle explain the $10^{-6}\,{\rm G}$ fields observed at cluster scales. 

On the other hand, the lower bound on magnetic fields in the intergalactic medium strongly points toward a primordial origin, since astrophysical processes connected to the presence of free charges and/or a medium have difficulties to operate in void regions among structures. Therefore, a primordial generation mechanism able to provide magnetic fields that fulfill the bound on the intergalactic magnetic field, and at the same time have high enough amplitude at large scales $\mathcal{O}(\rm Mpc)$ to initiate the galactic dynamo, can in principle explain all present observations of magnetic fields in the intergalactic medium, galaxies and clusters.

In \cite{Caprini:2014mja} it was shown that the model under analysis here can provide seeds larger than $10^{-23}\div 10^{-21}\,{\rm G}$ at $1\,{\rm Mpc}$, for a certain range of values of $n$, satisfying, at the same time, the lower bound on intergalactic magnetic fields.
Here we show that this statement holds true also when the strict constraints on scalar nongaussianities are taken into account, provided the auxiliary field does not stop rolling too far from the end of inflation.

More precisely, we know that for the set of inflationary parameters $H$ and $\xi$ for which the bound on $f_{\rm NL}^{\rm equil}$ is saturated (the top curve in figure \ref{plot:constraints}) the limit \eqref{bnv:constraintnB} is satisfied. We then proceed to investigate whether for these parameter values the intensity of the magnetic field can also be high enough to initiate the galactic dynamo. As done in \cite{Caprini:2014mja}, we therefore impose the fulfillment of the equality in equation \eqref{bnv:constraintnB}, obtaining a relation between the magnetic field intensity $B_{0}$ and correlation scale $L_{0}$, which, combined with the first equation in \eqref{eqs:inverse_cascade}, provides the value of $B_{0}$ and $L_{0}$ as function of $n_B$, actually independently on the generation mechanism. Using these values we calculate the amplitude of the magnetic field at the scale $\ell=1\,{\rm Mpc}$ as a function of $n$, using Eq.~\eqref{eq:nB} and the relation
\begin{equation}
B_{0}\left(\ell\right)=B_{0}\,\left(\frac{L_{0}}{\ell}\right)^{\frac{5-\left|2n+1\right|}{2}}\,.
\end{equation}

\begin{figure}
	\centering
	\includegraphics[width=0.7\textwidth]{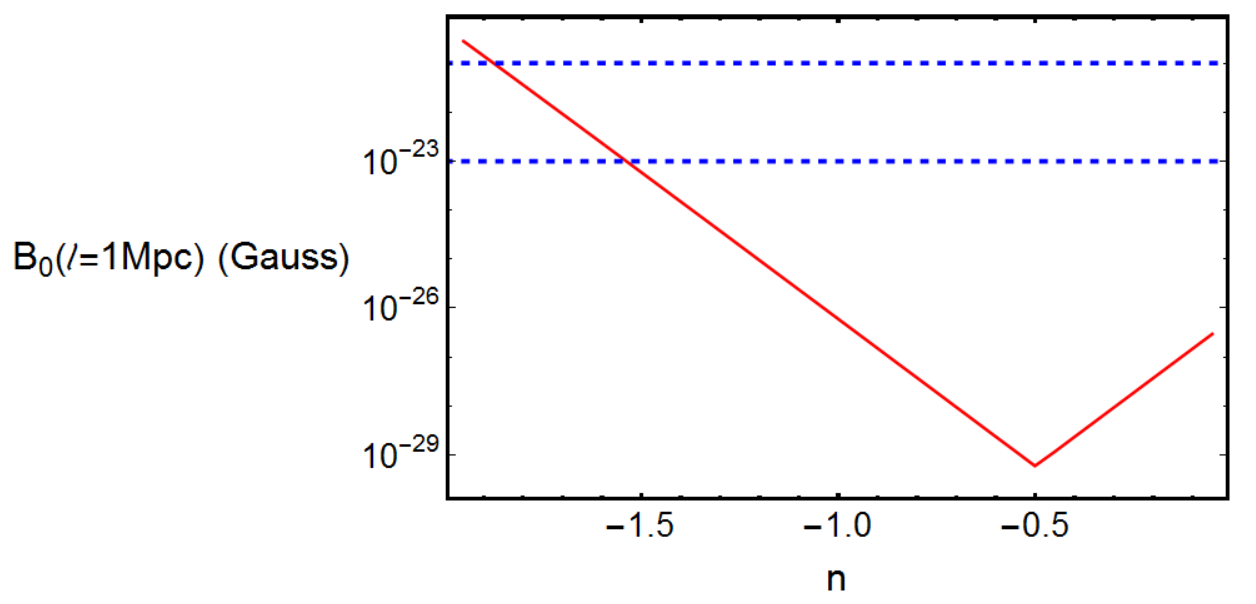}
	\caption{\small Magnetic field intensity at the Mpc scale as a function of $n$. In red, solid the value of $B_{0}$ at $1\,{\rm Mpc}$ obtained for the set of parameters $(H,n)$ which saturate the current upper bound on scalar non-gaussianities (the solid red curve in figure \ref{plot:constraints}). In blue, dashed, the estimations of the magnetic seed amplitude required to initiate the dynamo: $10^{-21}\,{\rm G}$ and $10^{-23}\,{\rm G}$ at $1\,{\rm Mpc}$.}
	\label{fig:galaxy}
\end{figure}

We compare the resulting $B_{0}\left(\ell=1\,{\rm Mpc}\right)$ with the required amplitude of magnetic seeds in galaxies on such scale, i.e. $10^{-23}\div 10^{-21}\,{\rm G}$. This is shown in figure \ref{fig:galaxy}. 

For $n=0$, i.e. for the case in which only the term $F_{\mu\nu}\tilde{F}^{\mu\nu}$ is active, the amplitude of the produced magnetic fields is too small to initiate the dynamo. On the other hand, for $n\leq -1.5$, inflationary magnetogenesis with added helicity can give rise to magnetic seeds at $1\,{\rm Mpc}$ higher than $10^{-23}\,{\rm G}$. The presence of the term $\sim F_{\mu\nu}F^{\mu\nu}$ is crucial at this aim, since it allows for redder spectral indexes $n_B<2$ than in the case where only the parity violating coupling is present. Imposing an intensity of $10^{-21}$~G at the Mpc scale, on the other hand, requires $n$ to be very close to $-2$. In this case the magnetic field is close to scale invariant and the behavior of the inverse cascade~\cite{Brandenburg:2016odr,Kahniashvili:2016bkp,Brandenburg:2018ptt} can lead to scalings different from those that gave eq.~(\ref{eqs:inverse_cascade}). Because of this, and of the possibility of additional infrared effects, the plot of Figure~\ref{fig:galaxy} might need corrections in this area.

\subsection{Inflationary model with the gauge field coupled to the inflaton.}
\label{sec:one:field}

We can also consider the scenario in which the gauge field $A_{\mu}$ is coupled to the inflaton, without the presence of an auxiliary field, i.e.:
\begin{equation}
	L=-\frac{1}{2}\nabla_{\mu}\varphi\nabla^{\mu}\varphi-V\left(\varphi\right)+I^{2}\left(\varphi\right)\left(-\frac{1}{4}F_{\mu\nu}F^{\mu\nu}+\frac{\gamma}{8}\epsilon_{\mu\nu\rho\lambda}F^{\mu\nu}F^{\rho\lambda}\right)\,.
\end{equation}
In this case, the non-minimal coupling between the inflaton and the gauge field leads to a source in the equation of motion of the inflaton perturbations which dominates with respect to the sources due to the gravitational coupling (see e.g. \cite{Barnaby:2012tk}). More precisely, the equation of motion for the inflaton $\varphi$ turns out to be
\begin{equation}
	\delta\varphi''_{\rm flat}+2\mathcal{H}\delta\varphi'_{\rm flat}-k^{2}\delta\varphi_{\rm flat}= II_{\varphi}\left[E_i*E_i+2\,\gamma\,\left(E_i*B_i\right)\right]+2\varphi'_{0}S^{\left(3\right)}+\frac{\varphi'_{0}}{\mathcal{H}}S'^{\left(3\right)}+\frac{\varphi'_{0}}{\mathcal{H}}S^{\left(2\right)}\,,
\end{equation}
where $S^{\left(3\right)}$ and $S^{\left(2\right)}$ have the same expressions as before, but $I=I\left(\varphi\right)$. Expression \eqref{R:approx} is still valid for this scenario. Differently from the model analysed previously, here the source is no more suppressed by the slow-roll parameter $\epsilon$. Consequently, in this case one finds that the constraints from scalar perturbations are too tight and do not allow for sufficient magnetogenesis.

\section{Summary and conclusions}

The mechanism of magnetogenesis with added helicity presented in~\cite{Caprini:2014mja} was able to account for the intergalactic magnetic fields inferred in~\cite{Neronov:1900zz,Taylor:2011bn,Vovk:2011aa} and initiate the galactic dynamo while keeping perturbation theory under control and with inflation happening at the (reasonably high) energies of the order of $10^6\div 10^{10}$~GeV. The work~\cite{Caprini:2014mja}, however, assumed that nongaussianities could be kept safely small by decoupling the scalar $\sigma$, whose slow roll excited the vacuum fluctuations of the gauge field, from the inflaton $\varphi$. Moreover, most of the analysis in that paper was performed under the assumption that the tensors sourced by the magnetic field during inflation would lead to a tensor-to-scalar ratio $r=0.2$, the value obtained by assuming that the BICEP2 signal~\cite{Ade:2014xna} was of primordial origin. 

In this work we have dropped the assumption that the sourced tensors lead to such a large value of $r$. Moreover, we have accounted for the fact, noted in~\cite{Ferreira:2014zia} (that appeared after~\cite{Caprini:2014mja}), that the fluctuations of the isocurvature field $\sigma$ sourced by the gauge field can oscillate into the observable curvature perturbation $\mathcal{R}$ if $\sigma$ rolls for a sufficient number of efoldings, possibly generating a very high level of non-gaussianity in the observed CMB spectrum. 

Here we have simultaneously evaluated the fluctuations in the inflaton originating from the oscillation from the $\sigma$ to the $\varphi$ sector and those directly induced by gravitational effects on the $\varphi$ sector. We have found that for all values of the parameter $n$ the former dominate over the latter even for $N_\sigma=1$. 

Our result is that, as long as the field $\sigma$ does not stop rolling until about one efold before the end of inflation, magnetogenesis with added helicity can account for the observed magnetic fields in the intergalactic medium and at the same time provide seeds high enough to initiate the galactic dynamo for a suitable range of values of the parameter $n$, while satisfying the observational bounds on equilateral nongaussianities, and with inflation occurring at $10^6\div 10^{8}$~GeV. If the bound on nongaussianities is saturated, the model then leads to chiral tensor modes with $r=10^{-6}\div 10^{-4}$. We can therefore conclude that the model of~\cite{Caprini:2014mja} is robust even once the stringent constraints from nongaussianities are accounted for, provided the above mentioned condition is met for the rolling of the auxiliary field\footnote{We have assumed, as it is often the case in works on inflationary magnetogenesis, that inflation is instantaneously followed by a radiation dominated phase. It would be interesting to study the more realistic situation where the inflaton performs many oscillations -- leading to a matter dominated Universe -- before decaying into radiation. The analysis of this system is nontrivial, but we generally expect that a long period of matter domination will make the magnetic field weaker.}.

It should be noted, however, that the constraints from CMB are only valid if the gauge field was already excited when CMB scales were leaving the horizon. Since the correlation length $L_0$ of the produced gauge field is well below the Mpc~\cite{Caprini:2014mja}, one can think of a scenario where $\sigma$ starts rolling after CMB scales, but before $L_0$, have left the horizon. In this (finely tuned) situation the constraints from CMB will not hold, and the mechanism can be effective even at higher energy scales for inflation.

\appendix

\section{Simplification of the gravitationally induced source terms for $\delta\varphi_{\rm flat}$}
\label{app:simpl_maxwell}

The equations of motion for the gauge field
\begin{align}
\partial^\mu\left[I^2\left(F_{\mu\nu}-\frac{\gamma}{2}\eta_{\mu\nu\rho\lambda}F^{\rho\lambda}\right)\right]=0
\end{align}
can be written in terms of the electric and magnetic fields defined in eq.~\eqref{eq:def_eb}, along with the Bianchi identities, as
\begin{align}
&\nabla\cdot {\bf E}=\nabla\cdot {\bf B}=0\,,\qquad {\bf B}'+2\,{\cal H}\,{\bf B}+\nabla\times{\bf E}=0\,,\nonumber\\
&{\bf E}'+2\frac{I'}{I}\,{\bf E}+2\,\frac{a'}{a}\,{\bf E}+2\,\gamma\,\frac{I'}{I}\,{\bf B}-\nabla\times {\bf B}=0
\end{align}

Now, in coordinate space,
\begin{align}
S^{(3)}=-\frac{I^2\,a^2}{2\,M_{pl}^2}\,\frac{\partial_j}{\nabla^2}\epsilon_{ijk}\,B_k\,E_i=\frac{I^2\,a^2}{2\,M_{pl}^2\,\nabla^2}\nabla\cdot\left({\bf E}\times{\bf B}\right)\,.
\end{align}
so that
\begin{align}
&2\,\varphi'_{0}\,S^{\left(3\right)}+\frac{\varphi'_{0}}{\mathcal{H}}\,S'^{\left(3\right)}=\frac{\varphi'_{0}}{\mathcal{H}\,a^2}\left(a^2\,S^{(3)}\right)'=-\frac{I^2\,\varphi'_{0}\,a^2}{2\,\mathcal{H}\,M_{pl}^2\,\nabla^2}\,\nabla\cdot\left[{\bf E}\times\left(\nabla\times{\bf E}\right)+{\bf B}\times\left(\nabla\times{\bf B}\right)\right]
\end{align}
with
\begin{align}
\nabla\cdot\left[{\bf E}\times\left(\nabla\times{\bf E}\right)\right](\bk)=k_i\int\frac{d\bq}{(2\pi)^{3/2}}E_i(\bk-\bq)\,k_j\,E_j(\bq)-k_i\int\frac{d\bq}{(2\pi)^{3/2}}(\bk-\bq)_i\,E_j(\bq)\,E_j(\bk-\bq)\,,
\end{align}
which can be simplified by noticing that 
\begin{align}
\int d\bq\,\bq_i\,E_j(\bq)\,E_j(\bk-\bq)=\int d\bp\,(\bk-\bp)_i\,E_j(\bk-\bp)\,E_j(\bp)
\end{align}
which implies
\begin{align}
\int d\bq\,\bq_i\,E_j(\bq)\,E_j(\bk-\bq)=\frac{\bk_i}{2}\int d\bq\,E_j(\bk-\bq)\,E_j(\bq)\,.
\end{align}

To sum up, we obtain
\begin{align}
\nabla\cdot\left[{\bf E}\times\left(\nabla\times{\bf E}\right)\right](\bk)=\bk_i\,\bk_j\int\frac{d\bq}{(2\pi)^{3/2}}E_i(\bk-\bq)\,E_j(\bq)-\frac{k^2}{2}\int\frac{d\bq}{(2\pi)^{3/2}}\,E_j(\bq)\,E_j(\bk-\bq)\,,
\end{align}
and, remembering that ${\bf B}$ gives contributions that are subdominant to ${\bf E}$, we get the approximate expression
\begin{align}
2\,\varphi'_{0}\,S^{\left(3\right)}+\frac{\varphi'_{0}}{\mathcal{H}}\,S'^{\left(3\right)}+\frac{\varphi'_{0}}{\mathcal{H}}\,S^{\left(2\right)}\simeq \frac{I^2\,\varphi'_{0}\,a^2}{2\,\mathcal{H}\,M_{pl}^2}\,\frac{\bk_i\,\bk_j}{k^2}\int\frac{d\bq}{(2\pi)^{3/2}}E_i(\bk-\bq)\,E_j(\bq)
\,,
\end{align}

\section*{Acknowledgments} M.C.G. thanks financial support from a Cariparo foundation grant and the APC (Astroparticules et Cosmologie) institute for hospitality during the development of this work. The work of L.S. is partially supported by the US NSF grant PHY-1520292. We thank the referees for having pointed out relevant conditions that need to be met to derive the result presented here.

\bibliographystyle{JHEP}
\bibliography{main}

\end{document}